%% file: main.tex
\documentclass[
		runningheads,
		10pt,
		dvipsnames
		]
		{llncs}
\makeatletter                   
\def\mdseries@tt{m}             
\makeatother                    
\input{packages}

\input{tikzinit}

\input{macros}
\input{notation}

\input{title}
\begin{document}
\maketitle
%
\input{abstract}

%
\input{intro}
\input{preliminaries}
\input{monitoring}
\input{replay}
\input{large}
\input{evaluation}
\input{related}
\input{conclusion}
\bibliographystyle{splncs03}
\bibliography{biblio}

\appendix
\input{appendix}

\end{document}

%% file: packages.tex
\usepackage{amsmath}
\usepackage{amssymb}
\usepackage{amsfonts}
\usepackage{times}
\usepackage{listings}
\usepackage{lscape}
\usepackage{todonotes}
\usepackage{algorithm, algpseudocode}
\usepackage[draft=true]{minted} 
\usepackage{subfig}
\usepackage{mathtools}
\usepackage{multirow}
\usepackage{booktabs} 
\usepackage[]{hyperref}
\hypersetup{
    bookmarks=true,         
    unicode=false,          
    pdftoolbar=true,        
    pdfmenubar=true,        
    pdffitwindow=false,     
    pdfstartview={FitH},    
    pdftitle={Bringing Runtime Verification Home},    
    pdfauthor={Antoine El-Hokayem and Ylies Falcone},     
    pdfsubject={Runtime/Formal Verification for Smart Apartments},   
    pdfkeywords={runtime verification, monitoring, formal verification, home, decentralized specifications}, 
    pdfnewwindow=true,      
    colorlinks=true,       
    linkcolor=red,          
    citecolor=violet,        
    filecolor=magenta,      
    urlcolor=cyan,           
    breaklinks=true,            
}

\usepackage{float}
\usepackage{wrapfig}

\let\subparagraph\paragraph
\usepackage[small,compact]{titlesec}
\usepackage[belowskip=-5pt,aboveskip=3pt]{caption}

\usepackage{geometry}

%% file: tikzinit.tex
\usepackage{tikz}
\usetikzlibrary{shapes,fadings,shadows,trees,matrix,intersections,patterns,decorations.markings,decorations.pathreplacing,arrows,automata,calc,decorations,fit,positioning,backgrounds}
\usetikzlibrary{shapes.geometric,decorations.pathmorphing}

\tikzset{
	aut/.style      = {auto,node distance=1cm,line width=2pt,>=to,thick,align=center,shorten >=1pt},
	location/.style = {state,fill=white!89!black,draw=white!50!black,minimum size=0.7cm,inner sep=0cm},
}

\tikzset{
	accept/.style  = {double},
	reject/.style  = {pattern=north east lines, pattern color=black!30},
	vtrue/.style  = {fill=green!60!black, text=white},
	vfalse/.style  = {fill=red!60!black, text=white, dotted, draw=black},
	vna/.style  = {fill=yellow!60!black, text=white},
	fit1/.style={black, rectangle, draw, solid, thin},
	smallarrow/.style={thin, -latex}
}

\newcommand{\tconnect}[5][]{ %
	\path[->,#1] (#2) edge [         ] node (#5)[label=#4]   {} (#3);%
}
\newcommand{\tconnectt}[6][]{ %
	\path[->,#1] (#2) edge [         ] node (#5)[label={[#6]:#4}]   {} (#3);%
}

%% file: macros.tex
\newcommand{\figref}[1]{Fig.~\ref{#1}}

\newcommand{\rtbl}[1]{Table~\ref{#1}}

\newcommand{\rsec}[1]{Sect.~\ref{#1}}

\newcommand{\rex}[1]{Ex.~\ref{#1}}

\newcommand{\squishlist}{
\begin{list}{-}
 { \setlength{\itemsep}{0pt}
    \setlength{\parsep}{1pt}
    \setlength{\topsep}{1pt}
    \setlength{\partopsep}{0pt}
    \setlength{\leftmargin}{0.6em}
    \setlength{\labelwidth}{1.5em}
    \setlength{\labelsep}{0.4em} } }
\newcommand{\squishend}{
 \end{list}  }

\newcommand{\wrt}[0]{wrt}

\newcommand{\cblock}[1]{}

\newcommand{\secref}[1]{Sect.~\ref{#1}}

\newcommand{\defas}[0]{\stackrel{{\scriptscriptstyle\rm def}}{=}}

%% file: notation.tex
\newcommand\mtt[1]{\ensuremath{\mathtt{#1}}}
\newcommand{\mrm}{\mathrm}

\newcommand\tuple[1]{\langle #1 \rangle}
\newcommand\setof[1]{\{#1\}}

\newcommand\vt{\ensuremath{\mathtt{true}}}
\renewcommand\vt{\textcolor{PineGreen}{\ensuremath{\top}}}
\newcommand\vf{\ensuremath{\mathtt{false}}}
\renewcommand\vf{\textcolor{Mahogany}{\ensuremath{\bot}}}

\newcommand\vna{\mathtt{\mathbf{?}}}

\DeclareMathOperator{\fdom}{dom}

\newcommand{\func}{\mathrm}

\newcommand{\ltlfw}[1]{\ensuremath{\Diamond_{\leq #1}}}
\newcommand{\ltlgw}[1]{\ensuremath{\square_{\leq #1}}}
\newcommand{\ltlx}[0]{\ensuremath{\bigcirc}}
\newcommand{\ltlu}[0]{\ensuremath{\, \mathrm{U}\, }}
\DeclareMathOperator{\ltlg}{\square}

\newcommand\verdict{\mathbb{B}_3}

\newcommand\aut{\mathcal{A}}

\newcommand\themis[0]{\texttt{THEMIS}}

\newcommand\mem{\mathcal{M}}

\newcommand\sys{\mathcal{I}}

\def\MemRep{\texttt{EHE}}

\newcommand\seval{\func{eval}}
\newcommand\sadv{\func{resolve}}
\newcommand\sadvs{\func{resolved}}

\newcommand\AP{\mathit{AP}}
\newcommand\APmons{{\mathit{AP}_{\mathrm{mons}}}}
\newcommand\smove{\mathrm{mov}}

\newcommand\MONID{\mathrm{m}}

\newcommand\comps{\mathcal{C}}

%% file: title.tex
\author{
	Antoine El-Hokayem\inst{1}
	\and
	Yli\`es Falcone\inst{2}
	\institute{
		Univ. Grenoble Alpes, CNRS, Grenoble INP, VERIMAG, 38000 Grenoble, France \and
		Univ. Grenoble Alpes, Inria, CNRS, Grenoble INP, LIG, 38000 Grenoble, France\\
		\email{firstname.lastname@univ-grenoble-alpes.fr}
	}
}

\title{%
	Bringing Runtime Verification Home%
	\thanks{%
		This work is supported by the French national program ``Programme Investissements d'Avenir IRT Nanoelec" (ANR-10-AIRT-05).
		The authors thank the Amiqual4Home (ANR-11-EQPX-0002) team, in particular S. Borkowski and J. Crowley for assisting in the case study and J. Cumin, for providing the collected data.
	}
}%

\subtitle{A Case Study on the Hierarchical Monitoring of Smart Homes\\ using Decentralized Specifications}

\titlerunning{Bringing Runtime Verification Home}
\authorrunning{Antoine El-Hokayem and Yli\`es Falcone}

%% file: abstract.tex
\begin{abstract}
	We use runtime verification (RV) to check various specifications in a smart apartment.
	The specifications can be broken down into three types: behavioral correctness of the apartment sensors, detection of specific user activities (known as activities of daily living), and composition of specifications of the previous types.
	The context of the smart apartment provides us with a complex system with a large number of components with two different hierarchies to group specifications and sensors: geographically within the same room, floor or globally in the apartment, and logically following the different types of specifications.
	We leverage a recent approach to decentralized RV of decentralized specifications, where monitors have their own specifications and communicate together to verify more general specifications.
	We leverage the hierarchies, modularity and re-use afforded by decentralized specifications to: (1) scale beyond existing centralized RV techniques, and (2) greatly reduce computation and communication costs.
\end{abstract}

%% file: intro.tex
%
Sensors and actuators are used to create ``smart'' environments which track the data across sensors and human-machine interaction.
One particular area of interest consists of homes (or apartments) equipped with a myriad of sensors and actuators, called \emph{smart homes}~\cite{crowley_smarthome}.
Smart homes are capable of providing added services to users.
These services rely on detecting the user behavior and the context of such activities~\cite{crowley_context}, typically detecting activities of daily living (ADL)~\cite{tapia_adl,chen_adl} from sensor information.
Detecting ADL allows to optimize resource consumption (such as electricity~\cite{home_energy}), improve the quality of life for the elderly~\cite{home_elderly} and users suffering from mild impairment~\cite{home_impaired}.

Relying on information from multiple sources and observing behavior is not just constrained to activities.
It is also used with techniques that verify the correct behavior of systems.
\emph{Runtime Verification} (RV)~\cite{HavelundG05,jlp/LeuckerS09,lncs/10457,Bartocci2017,BartocciFFR18} is a lightweight formal method which consists in verifying that a run of a system is correct {\wrt} a specification.
The specification formalizes the behavior of the system typically in logics (such as variants of Linear Temporal Logic, LTL~\cite{Pnueli77}) or finite-state machines.
Based on the provided specification, monitors are automatically synthesized to run alongside the system and verify whether or not the system execution complies with the specification.
RV techniques have been used for instance in the context of automotive~\cite{ex:autosar} and medical~\cite{ex:medical} systems.
In both cases, RV is used to verify communication patterns between components and their adherence to the architecture and their formal specifications.

While RV can be used to check that the devices in a smart home are performing as expected, we show it can be extended to monitor ADL, and complex behavior on the activities themselves.
We identify three classes of specifications for applying RV to a smart home.
The first class pertains to the system behavior.
These specifications are used to check the correct behavior of the sensors, and detect faulty sensors.
Ensuring that the system is behaving correctly is what is generally checked when performing RV.
However, it is also possible to use RV to verify other specifications.
The second class consists of specifications for detecting ADL, such as detecting when the user is cooking, showering or sleeping.
The third class pertains to user behavior.
These specifications can be seen as meta-specifications for both system correctness and ADL, they can include safety specifications such as ensuring that the user does not sleep while cooking, or ensuring that certain activities are only done under certain conditions.

However, standard RV techniques are not directly suitable to monitor the three classes of specifications.
This is mainly due to scalability issues arising from the large number of sensors, as typically RV techniques rely on a large formula to describe specifications.
Synthesizing centralized monitors from certain large formulas considered in this paper is not possible using the current tools.
Instead, we make use of RV with decentralized specifications~\cite{themis_issta,tosem}, as it allows monitors to reference other monitors in a hierarchical fashion.
The advantage of this is twofold.
First, it provides an abstraction layer to relate specifications to each other.
This allows specifications to be organized and changed without affecting other specifications, and even to be expressed with different specification languages.
Second, it leverages the structure and layout of the devices to organize the hierarchies.
On the one hand, we have a geographical hierarchy resulting from the spacial structure of the apartment from a given device, to a room, a floor, or the full apartment.
On the other hand, we have a logical hierarchy defined by the interdependence between specifications, i.e. ADL, specifications that use other ADL specifications, and specifications that combine sensor safety with ADL specifications.
For example, informally, consider checking two activities: sleeping and cooking, which can be expressed using formulae $\varphi_{\tt s}$ and $\varphi_{\tt c}$ respectively.
A monitor that checks whether the user is sleeping and cooking requires to check $\varphi_{\tt s} \land \varphi_{\tt c}$ and as such will replicate the monitoring logic of another monitor that checks $\varphi_{\tt s}$ alone, instead of re-using the output of that monitor.
The formula will be written twice, and changing the formula for detecting sleeping requires changing the formula for the monitor that checks both specifications.

Overall, we see our contributions as follows\footnote{An artifact~\cite{artifact} that contains data, documentation, and software, is provided to replicate and extend on the work.}:
\squishlist
\item
We apply decentralized RV to analyze traces of over 36,000 timestamps spanning 27 sensors in a real smart apartment (\secref{sec:house}).
\item
We show how to go beyond system properties, to specify ADL using RV, and more complex interdependent specifications defined on up to 27 atomic propositions (\secref{sec:propsgroups}).
\item
We leverage the hierarchies, modularity and re-use afforded by decentralized specifications (\secref{sec:monitoring}) to both be able to synthesize monitors and to reduce overhead when monitoring complex interdependent specifications (\secref{sec:decent-eval}).
\item We improve the existing data structures used for monitoring decentralized specifications, to account for large traces (\secref{sec:large}).
\item We use RV to effectively monitor ADL and identifying some insights and limitations inherent to using formal LTL specifications to determine user behavior (\secref{sec:decent-limitations}).
\item We elaborate on the advantages of modularity by adapting parts of the specification to the \emph{Activity Recognition with Ambient Sensing} (ARAS)~\cite{ARAS} dataset (\secref{sec:adaptation}).
\squishend

This paper extends existing work published in the proceedings of the the international conference on Runtime Verification (RV 2018)~\cite{El-HokayemF18a} with the following:
\squishlist
\item Providing a more detailed explanation of decentralized specifications and their dependency hierarchies (\secref{sec:decent});
\item Providing full details on trace generation, sensor polling, and trace replay using \themis{} (\secref{sec:themis});
\item Enhancing the existing data structures of~\cite{themis_issta} to support large traces, by elaborating on data structures, their operations, and strategies for garbage collection and lazy evaluation in \secref{sec:large};
\item Extending the evaluation section to include additional days where the trace is replayed, to illustrate changes in user behavior in \rsec{sec:decent-limitations}, adding more details for modifying the specification to improve precision and recall, and also illustrating adaptability to new environments by porting the specification to the ARAS dataset in \secref{sec:adaptation}.
\squishend

%% file: preliminaries.tex
\section{Writing Specifications for the Apartment}
\label{sec:prelem}
%
\subsection{Devices and Organization}
\label{sec:house}
%
We consider an actual apartment, with multiple rooms, where activities are logged using sensors.
Amiqual4Home~\cite{Paula} is an experimental platform consisting of a smart apartment, a rapid prototyping platform, and tools for observing human activity.
\paragraph{Overview of Amiqual4Home.}
The Amiqual4Home apartment is equipped with 219 sensors and actuators spread across 2 floors.
Amiqual4Home uses the OpenHab 6 integration platform for all the sensors and actuators installed.
Sensors communicate using the KNX, MQQT and UPnP protocols sending measurements to OpenHab over the local network, so as to preserve privacy.
The general layout of the apartment consists of 2 floors: the ground and first floors.
On the ground floor (resp. first floor), we have the following rooms: $\mtt{entrance}$, $\mtt{toilet}$, $\mtt{kitchen}$, and $\mtt{livingroom}$ (resp. $\mtt{office}$, $\mtt{bedroom}$, and $\mtt{bathroom}$).
Between the two floors, there is a connecting $\mtt{staircase}$.
This layout reveals a tree-like geographical hierarchy of components, where we can see the rooms at the leaves, grouped by floors then the whole apartment.
While in effect all device data is fed to a central observation point, it is reasonable to consider the hierarchy in the apartment as a simpler model to consider hierarchies in general, as one is bound to encounter a hierarchy at a higher level (from houses, to neighborhoods, to smart cities, etc.).
Furthermore, hierarchies appear when integrating different providers for devices in the same house.
\paragraph{Reusing the Orange4Home dataset.}
Amiqual4Home has been used to generate multiple datasets that record all sensor data, this includes an ADL recognition dataset~\cite{Paula} (ContextAct@A4H), and an energy consumption dataset~\cite{orangehome} (Orange4Home).
%
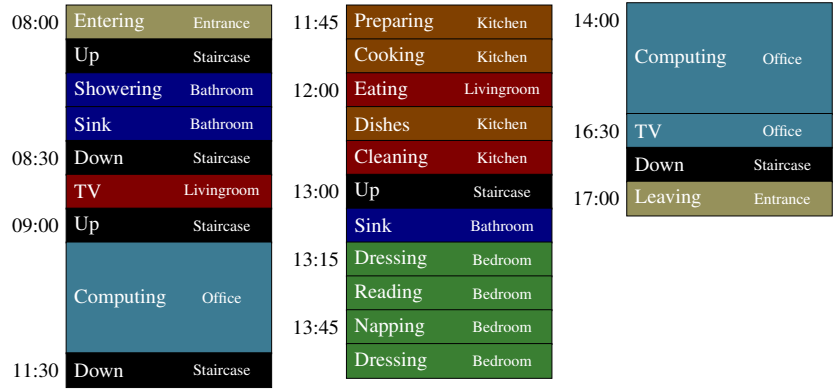
\begin{figure}[t]
	\centering
	\scalebox{0.85}{\input{tikz/schedule-july31}}
	\caption{Suggested Schedule (Tuesday, Jan 31 2017)}
	\label{fig:adl-proposed}
\end{figure}
%
In this paper, we reuse the dataset from~\cite{orangehome}.
The case study involved a person living in the apartment and following (loosely) a schedule of activities spread out across the various rooms.
The schedule was set out by the authors of~\cite{orangehome}.
Figure~\ref{fig:adl-proposed} displays the suggested schedule of activities for Tuesday, Jan 31 2017.
This allows us to nicely reconstruct the schedule from the result of monitoring the sensors.
Furthermore, the person living in the home provided manual annotations of the activities done, which helps us assess our specifications.
We chose to use the Orange4Home dataset over the ContextAct@A4H one as it involves only one person living in the house at a time which simplifies specifying and validating specifications.

\paragraph{Monitoring environment.}
In total, we formalize 22 specifications that make use of up to 27 sensors, and evaluate them over the course of a full day of activity in the apartment.
That is, we monitor the house (by replaying the trace) from 07:30 to 17:30 on a given day, by polling the sensors every 1 second, creating a trace of a total of 36,000 timestamps.
Specifications are elaborated in \secref{sec:propsgroups} and expressed as decentralized specifications~\cite{themis_issta} (recalled in \secref{sec:decent}).
Traces are replayed using the \texttt{THEMIS} tool~\cite{themis_tool} which supports decentralized specifications and provides a wide range of metrics.
We elaborate on the trace replay in \secref{sec:themis}.
%
\subsection{Property Groups} \label{sec:propsgroups}
\begin{figure}[t]
	\centering
	\includegraphics[width=12cm]{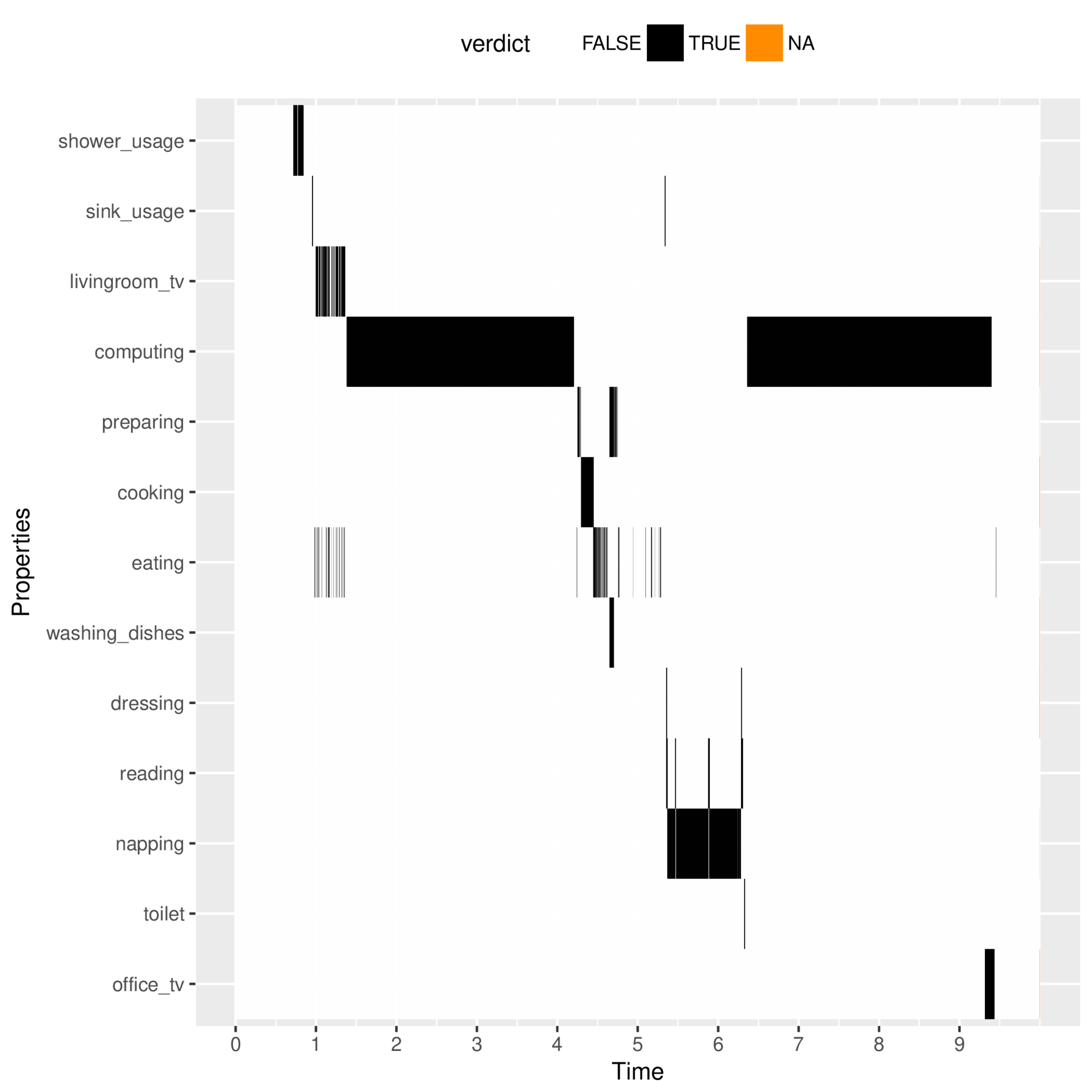}
	\caption{Detected ADL for Tuesday, Jan 31 2017. Time is in hours starting from 7:30.}
	\label{fig:adl-detected}
\end{figure}

%
We now express the specifications that describe different behaviors of components in the smart apartment.
Specifications can be subdivided into 3 groups: system-behavior specifications, user-behavior specifications, and meta-specifications on both system and user behavior.
The considered specifications are listed in Table~\ref{tbl:props-list}.
\begin{table}[t]
	\centering
	\caption{%
		Specifications considered in this paper. (*) indicates added ADL specifications.
		G indicates specification group: system (S), ADL (A), and meta-specifications (M).
		$|\mathrm{AP}|^\mathrm{d}$ (resp. ($|\mathrm{AP}|^\mathrm{c}$): atomic propositions needed to specify specification in decentralized (resp. centralized) specifications.
		$\mathrm{d}$ is the maximum depth of monitor dependencies.
	}
	\label{tbl:props-list}

	\begin{tabular}{l l l l c c c}\toprule

		\textbf{G} & \textbf{Scope} & \textbf{Name}           & \textbf{Description}                           & \textbf{$|\mathrm{AP}|^\mathrm{d}$} & \textbf{$|\mathrm{AP}|^\mathrm{c}$} &

		\textbf{$\mathrm{d}$}                                                                                                                                                                  \\

		\midrule

		S          & Room           & $\mtt{sc\_light}(i)$    & light switch turns on light ($i \in [0..3]$).  & 2                                   & 2                                   & 1 \\

		M          & House          & $\mtt{sc\_ok}$          & All light switches are ok.                     & 4                                   & 8                                   & 2 \\

		\midrule

		A          & Toilet         & $\mtt{toilet}^*$        & Toilet is being used.                          & 1                                   & 1                                   & 0 \\

		A          & Bathroom       & $\mtt{sink\_usage}$     & Sink is being used.                            & 1                                   & 2                                   & 1 \\

		A          & Bathroom       & $\mtt{shower\_usage}$   & Shower is being used.                          & 1                                   & 2                                   & 1 \\

		A          & Bedroom        & $\mtt{napping}$         & Tenant is sleeping on the bed.                 & 1                                   & 1                                   & 1 \\

		A          & Bedroom        & $\mtt{dressing}$        & Tenant is dressing, using the  closet.         & 2                                   & 3                                   & 1 \\

		A          & Bedroom        & $\mtt{reading}$         & Tenant is reading.                             & 3                                   & 5                                   & 2 \\

		A          & Office         & $\mtt{office\_tv}$      & Tenant is watching TV.                         & 1                                   & 1                                   & 1 \\

		A          & Office         & $\mtt{computing}$       & Tenant is using the computer.                  & 1                                   & 1                                   & 1 \\

		A          & Kitchen        & $\mtt{cooking}$         & Tenant is cooking food.                        & 2                                   & 2                                   & 1 \\

		A          & Kitchen        & $\mtt{washing\_dishes}$ & Tenant is cleaning dishes.                     & 2                                   & 3                                   & 1 \\

		A          & Kitchen        & $\mtt{kactivity}^*$     & Using cupboards and fridge.                    & 4                                   & 9                                   & 1 \\

		A          & Kitchen        & $\mtt{preparing}$       & Tenant is preparing to cook food.              & 2                                   & 11                                  & 2 \\

		A          & Living         & $\mtt{livingroom\_tv}$  & Tenant is watching TV.                         & 2                                   & 2                                   & 1 \\

		A          & Floor 0        & $\mtt{eating}$          & Tenant is eating on the table.                 & 2                                   & 2                                   & 1 \\

		\midrule

		M          & Floor 0        & $\mtt{actfloor}(0)$     & Activity triggered on floor 0.                 & 6                                   & 16                                  & 3 \\

		M          & Floor 1        & $\mtt{actfloor}(1)$     & Activity triggered on floor 1.                 & 7                                   & 11                                  & 3 \\

		M          & House          & $\mtt{acthouse}$        & Activity triggered in house                    & 2                                   & 27                                  & 4 \\

		M          & House          & $\mtt{notwopeople}$     & {No 2 simultaneous activities on different floors.} & 2                                   & 27                                  & 4 \\

		M          & House          & $\mtt{restricttv}$      & {No watching TV for more than 10s.}            & 2                                   & 3                                   & 3 \\

		M          & House          & $\mtt{firehazard}$      & {No cooking while sleeping.}                   & 2                                   & 3                                   & 2 \\

		\bottomrule
	\end{tabular}

\end{table}
\paragraph{System behavior.}
The first group of specifications consists in ensuring that the system behaves as expected.
That is, verifying that the sensors are working properly.
These properties are the subject of classical RV techniques~\cite{RVTutorial,LTL3Tools} applied to systems.
For the scope of this case study, we verify light switches as system properties.
We verify that for a given room $i$, whenever the switch is toggled, then the light must turn on until the switch is turned off.
We verify the property at two scopes, for a given room, and the entire apartment.
While this property appears simple to check, it does highlight issues with existing centralized techniques applied in a hierarchical way.
We develop the property in \secref{sec:monitors}, and show the issues in~\secref{sec:decent}.
\paragraph{ADL.}
The second group of specifications is concerned with defining the behavior of the user inferred from sensors.
The sensors available in the apartment provide us with a wealth of information to determine the user activities.
The list of activities of interest is detailed in~\cite{Katz-ADL} and includes activities such as cooking and sleeping.
By correctly identifying activities, it is possible to decide when to interact with the user in a smart setting~\cite{home_energy}, provide custom care such as nursing for the elderly~\cite{home_elderly}, or help users who suffer from mild impairment~\cite{home_impaired}.
Inferring activities done by the user is an interesting problem typically addressed through either data-based or knowledge-based methods~\cite{chen_adl}.
The first method consists in learning activity models from preexisting large-scale datasets of users’ behaviors by utilizing data mining and machine learning techniques.
The built models are probabilistic or statistical activity models such as Hidden Markov Model (HMM) or Bayesian networks, followed by training and learning processes.
Data-driven approaches are capable of handling uncertainty, while often requiring large annotated datasets for training and learning.
The second method consists in exploiting prior knowledge in the domain of interest to construct activity models directly using formal logical reasoning, formal models, and representation.
Knowledge-driven approaches are semantically clear, but are typically poor at handling uncertainty and temporal information~\cite{chen_adl}.
We elaborate on such limitations in \secref{sec:decent-limitations}.
Writing specifications can be seen as a knowledge-based approach to describe the behavior of sensors.
As such, we believe that runtime verification is useful to describe an activity as a specification over sensor outputs.
We formalize a specification for the following ADL activities described in~\cite{orangehome} (see Table~\ref{tbl:props-list}).
We re-use the traces to verify that our detected activities are indeed in line with the proposed schedule.
Figure~\ref{fig:adl-detected} displays the reconstructed schedule after detecting ADL with runtime verification.
Each specification is represented by a monitor that outputs (with some delay) for every timestamp (second) verdicts $\vt$ or $\vf$.
To do this, the monitor finds the verdict for a timestamp $t$ then respawns to monitor $t+1$.
Verdict $\vt$ indicates that the specification holds, that is, the activity is being performed.
The reconstructed schedule shows the eventual outcome of a specification for a given timestamp ignoring delay.
In reality some delay happens based on the specification itself, and the dependencies on other monitors.
\paragraph{Meta-specifications.}
Specifications of the last group are defined on top of the other specifications.
That is, we refer to a meta-specification as a specification that defines the interactions between various specifications.
While one can easily define specifications by defining predicates over existing ones, such as checking that the light switch specification holds in all rooms or whether or not detecting an activity was performed on a specific floor or globally in the house, we are more interested in specifications that relate to each other.
We consider a meta-specification that reduces fire hazards in the house.
In this case, we specify that the tenant should not cook and sleep at the same time, as this increases the risk of fire.
In addition to mutually excluding specifications, we can also constrain the behavior of existing specifications.
For example, we can specify a specification regulating the duration of watching TV to be at most $10$ timestamps.

%% file: tikz/schedule-july31.tex
\newcommand{\bsizea}{1.4cm}

\newcommand{\bblock}[6]{
  \node[aname, below=0 of a#1,#6](a#2) {\parbox{\bsizea}{#3}};
  \node[aloc, right=0 of a#2] (l#2) {#4};
  \begin{scope}[on background layer]
    \node[fit1, fit=(a#2)(l#2), #5] (f#2) {};
  \end{scope}
}
\newcommand{\bblockhead}[5]{
  \node[aname,#1](a#2) {\parbox{\bsizea}{#3}};
  \node[aloc, right=0 of a#2] (l#2) {#4};
  \begin{scope}[on background layer]
    \node[fit1, fit=(a#2)(l#2), #5] (f#2) {};
  \end{scope}
}
\begin{tikzpicture}[aut,
  fit1/.style={black, rectangle, draw, solid, thin,inner sep=0.2pt},
  slot/.style={rectangle, draw},
  smallarrow/.style={thin, -latex},
  aname/.style={font=\small, minimum width=\bsizea, anchor=west, minimum height=0.5cm, text=white},
  aloc/.style={yshift=0.0cm, minimum width=1.5cm,font=\scriptsize, text=white},
  atime/.style={minimum width=0.4cm},
  staircase/.style={fill=black},
  door/.style={fill=yellow!50!black},
  bath/.style={fill=blue!50!black},
  living/.style={fill=red!50!black},
  office/.style={fill=cyan!50!black},
  kitchen/.style={fill=orange!50!black},
  bed/.style={fill=OliveGreen},
  ]

	\node[aname](a0) at (0,0) {\parbox{\bsizea}{Entering}};
  \node[aloc, right=0 of a0] (l0) {Entrance};
  \begin{scope}[on background layer]
    \node[fit1, fit=(a0)(l0),door] (f0) {};
  \end{scope}

  \bblock{0}{1}{Up}{Staircase}{staircase}{}
  \bblock{1}{2}{Showering}{Bathroom}{bath}{}
  \bblock{2}{3}{Sink}{Bathroom}{bath}{}
  \bblock{3}{4}{Down}{Staircase}{staircase}{}
  \bblock{4}{5}{TV}{Livingroom}{living}{}
  \bblock{5}{6}{Up}{Staircase}{staircase}{}
  \bblock{6}{7}{Computing}{Office}{office}{minimum height=1.7cm}
  \bblock{7}{8}{Down}{Staircase}{staircase}{}

  \bblockhead{right=2.7 of a0}{9}{Preparing}{Kitchen}{kitchen}
  \bblock{9}{10}{Cooking}{Kitchen}{kitchen}{}
  \bblock{10}{11}{Eating}{Livingroom}{living}{}


  \bblock{11}{12}{Dishes}{Kitchen}{kitchen}{}
  \bblock{12}{13}{Cleaning}{Kitchen}{living}{}
  \bblock{13}{14}{Up}{Staircase}{staircase}{}
  \bblock{14}{15}{Sink}{Bathroom}{bath}{}
  \bblock{15}{16}{Dressing}{Bedroom}{bed}{}
  \bblock{16}{17}{Reading}{Bedroom}{bed}{}
  \bblock{17}{18}{Napping}{Bedroom}{bed}{}
  \bblock{18}{19}{Dressing}{Bedroom}{bed}{}

  \bblockhead{right=2.7 of a9, minimum height=1.7cm, anchor=north west, yshift=0.3cm}{20}{Computing}{Office}{office}
  \bblock{20}{21}{TV}{Office}{office}{}
  \bblock{21}{22}{Down}{Staircase}{staircase}{}
  \bblock{22}{23}{Leaving}{Entrance}{door}{}

  \node[left=0 of a0, atime]{08:00};

  \node[left=0 of a4, atime]{08:30};

  \node[left=0 of a6, atime]{09:00};

  \node[left=0 of a8, atime]{11:30};

  \node[left=0 of a9, atime]{11:45};

  \node[left=0 of a11, atime]{12:00};

  \node[left=0 of a14, atime]{13:00};
  \node[left=0 of a16, atime]{13:15};

  \node[left=0 of a18, atime]{13:45};

  \node[left=0 of a20, atime, yshift=0.6cm]{14:00};

  \node[left=0 of a21, atime]{16:30};

  \node[left=0 of a23, atime]{17:00};
\end{tikzpicture}

%% file: monitoring.tex
\section{Monitoring the Apartment}
\label{sec:monitoring}
%
%
We show how we monitor the apartment using decentralized specifications, while highlighting their advantages.
%
%
\subsection{Monitor Implementation}
\label{sec:monitors}
%
%
To monitor the apartment, we use LTL3 monitors~\cite{LTL3Tools}.
LTL3~\cite{0002LS07,BauerLS10} is a variant of the standard Linear Temporal Logic (LTL)~\cite{Pnueli77} giving a semantics to finite traces.
An LTL3 monitor is a complete and deterministic Moore automaton where states are labeled with the verdicts in a domain $\verdict = \setof{\vt, \vf, \vna}$.
Verdicts $\vt$ and $\vf$ respectively indicate that the current execution complies and does not comply with the specification, while verdict $\vna$ indicates that the verdict has not been determined yet.
Verdicts $\vt$ and $\vf$ are called final, as once the monitor outputs $\vt$ or $\vf$ for a given trace, it cannot output a different verdict for any suffix of that trace.
Using LTL3 monitors for representing specifications allows us to take advantage of the multiple RV tools that convert different specification languages to LTL3 monitors.
For our monitoring, we use the \texttt{THEMIS} tool~\cite{themis_tool} which is able to use both \texttt{ltl2mon}~\cite{LTL3Tools} and \texttt{LamaConv}~\cite{LamaConv} to generate monitors.
\texttt{ltl2mon} generates LTL3 monitors from LTL formulae, while \texttt{LamaConv} supports a wider range of languages such as Regular Expressions, Omega Regular Expressions, LTL,  LTL with past (pLTL),  Regular LTL (RLTL) and RLTL with past (pRLTL), and Structured Assertion Language for Temporal Logic (SALT)~\cite{SALTNFM}.
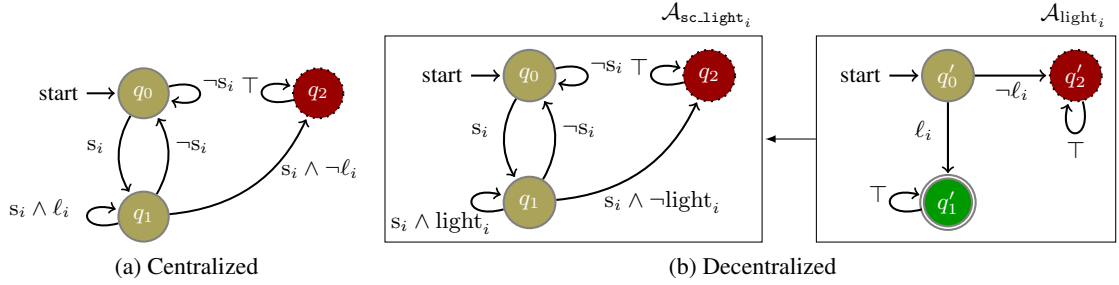
\begin{figure}[t]
	\centering
	\subfloat[Centralized]{
		\centering
		\resizebox{0.3\linewidth}{!}{\input{tikz/lightswitch-centralized}}
		\label{fig:sclight-cent}
	}
	\subfloat[Decentralized]{
		\centering
		\resizebox{0.6\linewidth}{!}{\input{tikz/lightswitch-decent}}
		\label{fig:sclight}
	}
	\caption{
		Monitor(s) for $\mtt{sc\_light}(i)$, for a given room $i$ in the house.
		The verdicts associated with the states are $\vf$: dotted red , $\vt$: double green, and $\vna$: single yellow.}
\end{figure}
\begin{example}[Check light switch]
\label{ex:sclight-cent}
	Let us consider property  $\mtt{sc\_light}(i)$ (sensor check light): ``Whenever a light switch is triggered in a room $i$ at some timestamp $t$, then the light must turn on at $t+1$ until the switch is turned off again''.
	Figure~\ref{fig:sclight-cent} shows the Moore automaton that represents the property.
	Starting from $q_0$ with verdict $\vna$, the automaton verifies that the property is falsified (as it is a safety property).
	That is, upon reaching $q_2$ the verdict will be $\vf$ for all possible extensions of a trace.
	%

\end{example}
For the scope of this paper and for clarity, we use LTL extended with two (syntactic) operators, mostly to strengthen and relax time constraints.
We consider the operator \emph{eventually within $t$} ($\ltlfw{t}$) which considers a disjunction of next operators.
It is defined as:
$\ltlfw{t} \mathit{ap} \defas \mathit{ap} \lor \ltlx \mathit{ap} \lor \ltlx\ltlx \mathit{ap} \lor ... \ltlx^t \mathit{ap}$, where $\mathit{ap}$ is an atomic proposition.
Intuitively, the eventually within states that $\mathit{ap}$ holds within a given number of timestamps.
Operator  $\ltlfw{t}$ allows us to relax the time constraints for a given atomic proposition.
Similarly, we consider the operator \emph{globally within $t$} ($\ltlgw{t}$) which is the dual of the previous operator: $\ltlgw{t} \mathit{ap} \defas \mathit{ap} \land \ltlx \mathit{ap} \land \ltlx\ltlx \mathit{ap} \land \ltlx^t \mathit{ap}$.
\begin{example}[Check light switch modalities]
\label{ex:sclight-ltl}
	The property expressed in \rex{ex:sclight-cent} can be expressed in LTL as:
	$\mtt{sc\_light}(i) \defas \ltlg(\mathrm{s}_i \implies \ltlx(\mathrm{\ell}_i \ltlu \neg\mathrm{s}_i))$.
	The property can be modified with the extra operators relax or constrain the time on the light.
	The relaxed property $\mtt{sc\_light}'(i) \defas \ltlg(\mathrm{s}_i \implies \ltlfw{3}(\mathrm{\ell}_i \ltlu \neg\mathrm{s}_i))$  allows the right-hand side of the implication to hold within any of the next 3 timestamps instead of immediately after.
	The bounded property $\mtt{sc\_light}''(i) \defas \ltlg(\mathrm{s}_i \implies \ltlgw{3}(\mathrm{\ell}_i))$ states that the light is on starting from the timestamp the switch is turned on and the subsequent two (for a total of 3).
	An example of such a property is the restriction on watching TV for a specific duration (Table~\ref{tbl:props-list}) where $\mtt{restricttv} \defas  \ltlg(\mtt{tv} \implies \ltlfw{10} \neg\mtt{tv})$.
\end{example}
%
\subsection{Decentralized Specifications} \label{sec:decent}
%
While simple specifications can be  expressed with both LTL and automata, it quickly becomes a problem to scale the formulae or account for hierarchies (see Sect.~\ref{sec:decent-advantages}).
As such, we use decentralized specifications~\cite{themis_issta}.
\paragraph{Overview.}
	Decentralized specifications consider a system of multiple components $\comps = \setof{\comps_1 \hdots \comps_n}$, where the set of all atomic propositions (noted $\AP$) (i) has a partition over all components, i.e., $\AP = \AP_1 \cup \hdots \cup \AP_n$ such that $\forall i,j \in [1 .. n], i \neq j \implies \AP_i \cap \AP_j = \emptyset$, and (ii) each component has at least one atomic proposition to monitor (i.e., $\forall i \in [1 .. n], \AP_i \neq \emptyset$).
	Details for assigning sensor information as atomic propositions for this case study are presented in \rsec{sec:trace-generation}.
	Furthermore, we have a set of monitor labels $\APmons$ (called \emph{monitor references}), that associates each monitor with a label.
	For this case study, each specification in \rtbl{tbl:props-list} is assigned a monitor labeled by its name.
	Each monitor $\aut_\mathrm{lbl}$ ($\mathrm{lbl} \in \APmons$) is a Moore automaton (detailed in \rsec{sec:monitors}) and is assigned to a single component.
	A monitor $\aut_\mathrm{lbl}$ assigned to component $\comps_j \in \comps$ utilizes the alphabet $\AP_\mathrm{lbl} = \AP_j \cup (\APmons \setminus \setof{\mathrm{lbl}})$.
	That is, it contains the atomic propositions local to the component (in $\AP_j$), and the references to all dependent monitors excluding itself ($\APmons \setminus \setof{\mathrm{lbl}}$).
A decentralized trace is a partial function that assigns each component and timestamp with an event.
A monitor reference is evaluated as if it were an oracle. That is, to evaluate a monitor reference $\mathrm{lbl}$ at a timestamp $t$, the monitor referenced ($\aut_\mathrm{lbl}$) is executed starting from the initial state on the trace starting at $t$.
The atomic proposition $\mathrm{lbl}$ at $t$ takes the value of the final verdict reached by the monitor.
\begin{example}[Decentralized light switch]
	Figure~\ref{fig:sclight} shows the decentralized specification for the check light property from \rex{ex:sclight-cent}.
	We have two monitors $\aut_{\mtt{sc\_light}_i}$ and $\aut_{\mrm{light}_i}$.
	They are respectively attached to the light switch and light bulb components.
	In the former, the atomic propositions are either related to observations on the component ($\mrm{s}_i$, switch on), or references to other monitors (${\mrm{light}_i}$).
	The light switch monitor first waits for the switch to be on to reach $q_1$.
	In $q_1$, at some timestamp $t$, it needs to evaluate reference ${\mrm{light}_i}$ by running the trace starting from $t$ on monitor $\aut_{\mrm{light}_i}$.
	Monitor $\aut_{\mrm{light}_i}$ then reads the value of $\mrm{\ell}_i$ at $t$ from the trace, and moves to $q'_1$ or $q'_2$ depending on its value, and sends the verdict $\vt$ or $\vf$ respectively back to monitor $\aut_{\mtt{sc\_light}_i}$.
	The returned verdict is associated with the reference ${\mrm{light}_i}$ for timestamp $t$ allowing monitor $\aut_{\mtt{sc\_light}_i}$  to evaluate its own transition at $t$.
\end{example}
\paragraph{Assumptions.}
The assumptions of decentralized specifications on the system are as follows:  no monitors send messages that contain wrong information;
no messages are lost, they are eventually delivered in their entirety but possibly out-of-order; all components share one logical discrete clock marked by round numbers indicating relevant transitions in the system specification.
While security is a concern in the smart apartment setting, the first two assumptions are met in this case study as the apartment sensor network operates on the local network, and we expect monitors to be deployed by the sensor providers, and users of the apartment.
The last assumption is also met in the smart setting, as all sensors share a global clock.
\paragraph{Hierarchical dependencies.}
Decentralized specifications allow us to analyze the dependencies between various monitors, and organize them in logical hierarchies represented as directed acyclic graphs (DAGs).
The DAGs help us relate specifications to other specifications and analyze the inter-dependent behavior of monitors.
We elaborate on the benefits of the hierarchical dependencies in \secref{sec:decent-advantages}.
\begin{example}[Hierarchical dependencies]
	Figure~\ref{fig:dependencyDAG} presents the dependency DAG of specification $\mtt{preparing}$.
	We can see that specification $\mtt{preparing}$ depends directly on both specifications $\mtt{kactivity}$ and $\mtt{cooking}$.
	Specification $\mtt{kactivity}$ depends on specifications $\mtt{cubpoard}$, $\mtt{sink\_water}$, $\mtt{presence}$, and $\mtt{fridge\_door}$, as it depends on the tenant being present in the kitchen, opening or closing cupboards or the fridge, or using the sink.
	The later specifications do not depend on other specifications but on direct observations from the components.
	We note that while $\mtt{presence}$ is not used in this case study to determine the cooking activity, since a tenant can start cooking and leave the kitchen.
	One could imagine that specifications can share dependencies, as such the hierarchy is indeed best represented as a DAG.
	Let us consider the monitor checking specification $\mtt{cupboard}$.
	Since we have 5 cupboard doors, we have 5 sensors in total (1 for each door).
	The monitor observing the 5 different observations simply checks if one is open and relays its verdict upwards, transmitting only the summary of observations instead of the totality.
	In this example, the hierarchy can be seen starting from different sensors on the same component, and expanding geographically to the different components in the room (kitchen).
\end{example}
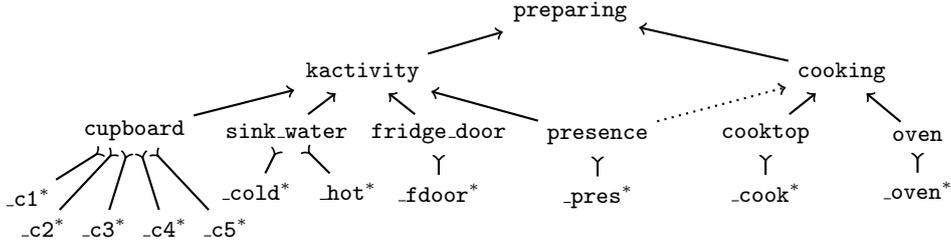
\begin{figure}[t]
	\centering
	\input{tikz/dep.tex}
	\caption{Dependencies for $\mtt{preparing}$. * indicates an atomic proposition of a component.}
	\label{fig:dependencyDAG}
\end{figure}
%
\subsection{Advantages of Decentralized Specifications}
\label{sec:decent-advantages}
%
\paragraph{Modularity and re-use.}
Monitor references in decentralized specifications allow specifications writers to modularize behavior.
Given that a monitor represents a specific specification, this same monitor can be re-used to define more complex specifications at a higher level, without consideration for the details needed for this specification.
This allows specification writers to reason at various levels about the system specification.

Let us consider the ADL specification $\mtt{cooking}$ (resp. $\mtt{sleeping}$) which specifies whether the tenant is cooking (resp. sleeping) in the apartment.
One can reason about the meta-specification $\mtt{firehazard}$ using both $\mtt{cooking}$ and $\mtt{sleeping}$ specifications without considering the lower level sensors that determine these specifications, that is:
\[
	\mtt{firehazard} \defas \ltlg(\mtt{sleeping} \implies \neg\mtt{cooking}).
\]
While we can define $\mtt{cooking}$ as:
\[
	\mtt{cooking} \defas  \mtt{kitchen\_presence} \land \ltlfw{5}(\mtt{kitchen\_cooktop} \lor \mtt{kitchen\_oven}).
\]
Additionally, any specification that requires either $\mtt{sleeping}$ or $\mtt{cooking}$ specifications can re-use the verdict outputted by their respective monitors.
For example, specifications $\mtt{actfloor}(0)$ and $\mtt{actfloor}(1)$ require the verdicts from monitors associated with $\mtt{cooking}$ and $\mtt{sleeping}$, respectively, since cooking happens on the ground floor while sleeping on the first floor.
Furthermore, we can disjoin $\mtt{actfloor}(0)$ and $\mtt{actfloor}(1)$ to easily specify that there is some activity in the house, $\mtt{acthouse} \defas  \mtt{actfloor}(0) \lor \mtt{actfloor}(1)$.
While specification $\mtt{acthouse}$ can be seen as a quantified version of $\mtt{actfloor}(i)$, we can use modular specifications for behavior, for example we can verify the triggering of an alarm in the house within 5 timestamps of detecting a fire hazard, i.e. $\mtt{checkalert} \defas  \mtt{firehazard} \implies \ltlfw{5}(\mtt{firealert})$.

In addition to providing a higher level of abstraction and reasoning about specifications, the modular structure of the specifications present three additional advantages.
\begin{enumerate}
	\item 
The first is that sub-specifications can change without affecting the meta-specifications, that is if the sub-specification $\mtt{cooking}$ is changed (possibly to account for different sensors), no changes need to be propagated to specifications $\mtt{firehazard}$, $\mtt{actfloor}(0)$, $\mtt{acthouse}$, and $\mtt{checkalert}$.
	\item 
The second advantage is controlling duplication of computation and communication, as such sensors do not have to send their observations constantly to all monitors that verify the various specifications.
Specification $\mtt{cooking}$ requires knowledge from the kitchen presence sensor, the kitchen cooktop (being enabled) and the kitchen oven.
Without any re-use these three sensors (presence, cooktop, and oven) need to send their information to monitors checking: $\mtt{firehazard}$, $\mtt{actfloor}(0)$, $\mtt{acthouse}$, and $\mtt{checkalert}$.
	\item 
The third advantage is a consequence of modeling explicitly the dependencies between specifications.
This allows the monitoring to take advantage of such dependencies and place the monitors that depend on each other closer depending on the hierarchy, either geographically (i.e., in the same room or floor) or logically (i.e., close to the monitors of the dependent sub-specifications).
Furthermore, knowing the explicit dependencies between specifications allows the user to choose a placement for their monitors, adjusting the placement to the system architecture.
In the case a placement is not possible, it is possible to create intermediate specifications that simply relay verdicts of other monitors, to transitively connect all components that are not connected.
\end{enumerate}
\paragraph{Abstraction from implementation.}
One setback for learning-based techniques to detect ADL is their specificity to the environment.
That is, the training set is specific to a house layout, user profile (i.e., elderly versus adults)~\cite{KasterenEK10}.

Decentralized specifications define modular specifications that can be composed together to form bigger and more complex specifications.
By using references to monitors, we leave the implementation of the specification to be specific for the house or user profile.
Using our existing example, $\mtt{cooking}$ is implemented based on the available sensors in the house, which would change for different houses.
However, the meta-specifications such as $\mtt{firehazard}$ can be defined independently from the implementation of both $\mtt{cooking}$ and $\mtt{sleeping}$.

Furthermore, using monitor references, which are treated as oracles, opens the door to utilizing existing techniques in the literature based on other formalisms (not based on automata).
That is, as a reference is expected to eventually evaluate to $\vt$ or $\vf$, any decision procedure can be incorporated to form more complex specifications.
For example, one can use the various machine learning techniques~\cite{crowley_context,KasterenEK10,tapia_adl} to define monitors that detect specific ADLs, then reference them in order to define more complex specifications.
\paragraph{Scalability.}
Decentralized specifications allow for a higher level of scalability when writing specifications, and also when monitoring.
By using decentralized specifications, we restrict the atomic propositions of monitors to (i) the local atomic propositions of the components they are attached to and (ii) references to other monitors (see \secref{sec:decent}).
This greatly reduces the number of atomic propositions to consider when synthesizing the monitor and reduces its size, as the sub-specifications are offloaded to another monitor.

For example, let us consider writing specifications using LTL formulae.
The classical algorithm that converts LTL to Moore automata is doubly exponential in the size of the formula counted in terms of atomic propositions (to form events)~\cite{LTL3Tools}.
Therefore, reducing both the size of the formula and the number of atomic propositions used in the formula helps significantly when synthesizing the monitors, allowing us to scale beyond the limits of existing tools.
For a large formula, and the larger formulas considered in this paper, it becomes impossible to generate a central monitor using the existing synthesis techniques.
Decentralized specifications provide a way to manage the large formula by subdividing it into subformulas.
The decomposition ensures that the formula evaluates to the same verdict given the same observations, at the cost of added delay.
\begin{example}[Synthesizing the check light monitor]
\label{ex:sclight-synthesis}
	Recall the system property $\mtt{sc\_light}(i)$ in \rex{ex:sclight-ltl} responsible for verifying that in a room $i$ a light switch does indeed turn a light bulb on until it is turned off.
	We recall the LTL specification
	$\mtt{sc\_light}(i) \defas \ltlg(\mathrm{s}_i \implies \ltlx(\mathrm{\ell}_i \ltlu \neg\mathrm{s}_i))$.
	To verify the property across $n$ rooms of the house, we formulate a property $\mtt{sc\_ok} \defas \bigwedge_{i \in [0 .. n]} \mtt{sc\_light}(i)$.
	In the case of a decentralized specification the formula will reference each monitor in each room, leading to a conjunction of at $n$ atomic propositions.
	However, in the case of a centralized specification, the specification needs to be written as: $\mtt{sc\_ok}^\mrm{cent} \defas \bigwedge_{i \in [0 \ldots n]} \ltlg(\mathrm{s}_i \implies \ltlx(\mathrm{\ell}_i \ltlu \neg\mathrm{s}_i))$, which is significantly more complex as a formula consisting of $4n$ operators (to cover the sub-specification), along $n$ conjunctions, and defined over each sensor and light bulb atomic propositions ($2n$).
	Given that monitor synthesis is doubly exponential, both \texttt{ltl2mon}~\cite{LTL3Tools} and \texttt{lamaconv}~\cite{LamaConv} require significant resources and time to generate the minimal Moore automaton (in our case\footnote{On an Intel(R) Core(TM) i7-6700HQ CPU, using 16GB RAM, and running openjdk 1.8.0\_172, with ltl2mon 0.0.7.}, both tools where unable to generate the monitor for $n = 3$ after an hour to timeout).
\end{example}

%% file: tikz/lightswitch-centralized.tex
\begin{tikzpicture}[aut]
	\node[location, vna, initial] (q0) at (0,0) {$q_0$};
	\node[location, vna, below=of q0] (q1) {$q_1$};
	\node[location, reject, vfalse, right=1.7 of q0] (q2) {$q_2$};

	\tconnectt[bend right]{q0}{q1}{$\mathrm{s}_i$}{t1}{left, xshift=-0.2cm}
	\tconnectt[bend right]{q1}{q0}{$\neg\mathrm{s}_i$}{t2}{right, xshift=0.1cm}
	\tconnectt[loop right]{q0}{q0}{$\neg\mathrm{s}_i$}{t2}{right, xshift=-0.2cm}
	\tconnectt[loop left]{q1}{q1}{$\mathrm{s}_i \land \mathrm{\ell}_i$}{t2}{left}
	\tconnectt[loop left]{q2}{q2}{$\top$}{t2}{left,xshift=0.2cm}
	\tconnectt[bend right]{q1}{q2}{$\mathrm{s}_i \land \neg\mathrm{\ell}_i$}{t2}{right, xshift=0.4cm}
\end{tikzpicture}

%% file: tikz/lightswitch-decent.tex
\begin{tikzpicture}[aut]
	\node[location, vna, initial] (q0) at (0,0) {$q_0$};
	\node[location, vna, below=of q0] (q1) {$q_1$};
	\node[location, reject, vfalse, right=1.7 of q0] (q2) {$q_2$};
	\node[left=1 of q0] (q0fit) {};

	\tconnectt[bend right]{q0}{q1}{$\mathrm{s}_i$}{t1}{left, xshift=-0.2cm}
	\tconnectt[bend right]{q1}{q0}{$\neg\mathrm{s}_i$}{t2}{right, xshift=0.1cm}
	\tconnectt[loop right]{q0}{q0}{$\neg\mathrm{s}_i$}{t3}{right, xshift=-0.2cm}
	\tconnectt[loop left]{q1}{q1}{$\mathrm{s}_i \land \mathrm{{light}}_i$}{t6}{left,xshift=0.5cm,yshift=-0.4cm}
	\tconnectt[loop left]{q2}{q2}{$\top$}{t4}{left,xshift=0.2cm}
	\tconnectt[bend right]{q1}{q2}{$\mathrm{s}_i \land \neg\mathrm{light}_i$}{t5}{below, xshift=0.4cm,yshift=-0.4cm}

	\node[location, vna, initial, right=2.5 of q2] (q00) {$q'_0$};
	\node[location, accept, vtrue, below=of q00] (q01) {$q'_1$};
	\node[location, reject, vfalse, right=of q00] (q02) {$q'_2$};
	\node[left=1 of q00] (q00fit) {};

	\tconnectt[]{q00}{q01}{$\mathrm{\ell}_i$}{t01}{left, xshift=-0.2cm}
	\tconnectt[]{q00}{q02}{$\neg\mathrm{\ell}_i$}{t01}{below, yshift=-0.2cm}

	\tconnectt[loop left]{q01}{q01}{$\top$}{t4}{left,xshift=0.2cm}
	\tconnectt[loop below]{q02}{q02}{$\top$}{t4}{below}

	\node (mswitch) [fit1, inner sep=5pt, inner xsep=10pt, fit = (q0fit) (q0) (q1) (q2)] {};
	\node (mlight) [fit1, inner sep=5pt, inner xsep=5pt, fit = (q00fit) (q00) (q01) (q02)] {};

	\path[smallarrow]
		(mlight) edge (mswitch);

	\node (labelswitch) [anchor=south east] at ($(mswitch.north east)$) {$\aut_{\mtt{sc\_light}_i}$};

	\node (labellight) [anchor=south east] at ($(mlight.north east)$) {$\aut_{\mrm{light}_i}$};
\end{tikzpicture}

%% file: tikz/dep.tex
\begin{tikzpicture}[aut]

    \node[] (preparing) at (0,0) {$\mtt{preparing}$};

    \node[below left=0.3 and 1 of preparing] (kactivity) {$\mtt{kactivity}$};
    \node[below right=0.3 and 2 of preparing] (cooking) {$\mtt{cooking}$};
    \tconnect{kactivity}{preparing}{}{r0}
    \tconnect{cooking}{preparing}{}{r1}


    \node[below=0.3 of kactivity, xshift=-1.0cm] (sink){$\mtt{sink\_water}$};
    \node[below=0.3 of kactivity, xshift=1.0cm] (fridge){$\mtt{fridge\_door}$};

    \node[right=0.3 of fridge, yshift=-0.05cm] (presence){$\mtt{presence}$};
    \node[left=0.3 of sink] (cupboard){$\mtt{cupboard}$};
    \tconnect{sink}{kactivity}{}{kact1}
    \tconnect{fridge}{kactivity}{}{kact2}
    \tconnect{presence}{kactivity}{}{kact3}
    \tconnect{cupboard}{kactivity}{}{kact4}
    \tconnect[dotted]{presence}{cooking}{}{kact5}

    \node[below=0.3 of cooking, xshift=-1cm] (cooktop){$\mtt{cooktop}$};
    \node[below=0.3 of cooking, xshift=1cm, yshift=-0.1cm] (oven){$\mtt{oven}$};
    \tconnect{cooktop}{cooking}{}{c1}
    \tconnect{oven}{cooking}{}{c2}

    \node[below=0.3 of oven](ooven){$\mtt{\_oven}^*$};
    \node[below=0.3 of cooktop](ocook){$\mtt{\_cook}^*$};
    \tconnect[-<]{ooven}{oven}{}{o01}
    \tconnect[-<]{ocook}{cooktop}{}{o02}

    \node[below=0.3 of presence](opres){$\mtt{\_pres}^*$};
    \tconnect[-<]{opres}{presence}{}{o03}

    \node[below=0.3 of fridge](ofdoor){$\mtt{\_fdoor}^*$};
    \tconnect[-<]{ofdoor}{fridge}{}{o04}

    \node[below=0.3 of sink,xshift=-0.4cm, yshift=-0.05cm](ocold){$\mtt{\_cold}^*$};
    \node[below=0.3 of sink,xshift=0.8cm, yshift=-0.05cm](ohot){$\mtt{\_hot}^*$};

    \tconnect[-<]{ocold}{sink}{}{o05}
    \tconnect[-<]{ohot}{sink}{}{o06}

    \node[below=0.8 of cupboard,xshift=-1.2cm](oc5){$\mtt{\_c2}^*$};
    \node[below=0.8 of cupboard,xshift=-0.4cm](oc2){$\mtt{\_c3}^*$};
    \node[below=0.8 of cupboard,xshift=0.4cm](oc3){$\mtt{\_c4}^*$};
    \node[below=0.8 of cupboard,xshift=1.2cm](oc4){$\mtt{\_c5}^*$};
    \node[below=0.4 of cupboard,xshift=-1.4cm](oc1){$\mtt{\_c1}^*$};

    \tconnect[-<]{oc1}{cupboard}{}{o07}
    \tconnect[-<]{oc2}{cupboard}{}{o08}
    \tconnect[-<]{oc3}{cupboard}{}{o09}
    \tconnect[-<]{oc4}{cupboard}{}{o10}
    \tconnect[-<]{oc5}{cupboard}{}{o10}


\end{tikzpicture}

%% file: replay.tex
\section{Trace Replay with \themis{}}
\label{sec:themis}
%
%
To perform monitoring we use \themis{}~\cite{themis_tool} which is a tool for defining, handling, and benchmarking decentralized specifications and their monitoring algorithms.
For replaying the trace, we perform monitoring by defining a start time, an end time and a polling interval.
For this case study, for a given date, we use 07:30 as start time, 17:30 as an end time, and a 1-second polling interval.

We first overview \themis{} in \secref{sec:themis-overview}.
Then, in \secref{sec:trace-generation}, we elaborate on the trace format provided in the public dataset, and our adaptation for replay to perform the monitoring.
In brief, the process consists of extracting each sensor data converting it to observations (atomic propositions and verdicts), and passing the observation to a logical component for multiple related sensors.
Later in \secref{sec:large}, we introduce extra considerations when monitoring large traces.
%
%
\subsection{\themis{}}
\label{sec:themis-overview}
%
%
\paragraph{Overview.}
\themis{}~\cite{themis_tool} is a tool to facilitate the design, development, and analysis of decentralized monitoring algorithms; developed using Java and AspectJ.
It consists of a library and command-line tools.
\themis{} provides an API, data structures, and measures for decentralized monitoring.
These building blocks can be reused or extended to modify existing algorithms, design new algorithms, and elaborate new approaches to assess existing algorithms.
\themis{} encompasses existing approaches~\cite{FAMTL,DecentMon} that focus on presenting one global formula of the system from which they derive multiple specifications, and in addition supports any decentralized specification~\cite{tosem}.

\paragraph{Monitoring.}
\themis{} defines two phases for a monitoring algorithm: setup and monitor.
In the first phase, the algorithm creates and initializes the monitors, connects them to each other so they can communicate, and attaches them to components so they receive the observations generated by components.
In the second phase, each monitor receives observations at a timestamp based on the component it is attached to.
The monitor can then perform some computation, communicate with other monitors, abort monitoring or report a verdict.
The two distinct phases separate the monitor generation (monitor synthesis) problem from the monitoring~\cite{themis_issta}, giving algorithms the freedom to generate monitors and deploy them on components, while integrating with existing tools for monitor synthesis such as~\cite{LTL3Tools,LamaConv}.
The monitors used in this case study use similar logic than \emph{choreography}~\cite{DecentMon}, as they are defined over a shared global clock.
All monitors start monitoring at $t = 0$.
A monitor checks the compliance of the specification for a given timestamp $t$, which could take a fixed delay $d$ to check.
After reaching the delay at $t+d$, the monitor reports the verdict for $t$ to all other monitors that depend on it, and starts monitoring the specification again for $t+1$ (i.e., it \emph{respawns}).
As such, the communication between monitors consists of sending verdicts for given timestamps.
%
%
\subsection{Generating the Trace}
\label{sec:trace-generation}
%
%
\paragraph{Provided trace.}
The trace from~\cite{orangehome} is given as a database with a table for each sensor.
We extract each table as a \emph{csv} file for each sensor.
The provided sensor data is stored as entries of values associated with timestamps, representing the changes in the sensor data across time.
Typically, a new entry is provided whenever a change in the sensor data occurs.
The provided data range over Boolean-like, integer, or real domains.
\paragraph{Generating atomic propositions.}
The sensor data needs to be processed to create observations, as LTL3 monitors (see \secref{sec:monitors}) operate on atomic propositions.
Each sensor is implemented as an input (\emph{Periphery} in \texttt{THEMIS}) to a logical component.
For example, for the shower water, we use both cold and hot water sensors but define only a single component (``shower water''), from an RV perspective, ``hot'' and ``cold'' are multiple observations passed to the ``shower water'' component.
To process different sensor data, we implemented two peripheries: \emph{SensorBool} and \emph{SensorThresh}.
The first periphery parses Boolean values from the \emph{csv} file associated with timestamps.
The processing assigns Boolean values $\vt$ (resp. $\vf$) based on sensor data such as: "ON" (resp. "OFF"), and "OPEN" (resp. "CLOSED").
The second periphery reads real (double) values, and returns a Boolean based on whether the number is below or above a certain threshold.
Both peripheries associate each atomic proposition with the generated Boolean to generate an observation.
\paragraph{Synchronizing traces.}
The provided dataset only provides sensor updates, that is, the data only contains timestamps and values for a sensor when the value changes.
Our monitoring strategy, however, requires polling the devices at given fixed time intervals.
Since the system has a global clock, to synchronize observations, our periphery implementations synchronize on a date at the start and an increase (in our case 1 second) and a default Boolean value for the observation.
When polled, the periphery returns the default value if nothing is observed yet, or the last value observed otherwise.
The last value observed is updated when changes occur in the \emph{csv} file.
In short, we interpolate values between changes to return the oldest value before a change.
\paragraph{Determining the polling rate.}
We leverage the global clock of the system to evaluate the specification synchronously for all components.
As such, we need a fixed interval to poll the monitors in order to evaluate the specification, that is, we take the necessary transition in each of the automata.
We refer to this interval as the \emph{polling rate}.
The polling rate determines the frequency of evaluation of the specification; the higher the rate, the more rounds, and the more monitors process and communicate.
To determine the minimal rate, we consider the rate of change for all sensors involved in the specification.
We are interested in ensuring that no sensor changes twice in between the evaluation of the specification.
To do so, we write a simple program that processes the trace files for each sensor in an input specification, to determine the rate of change.
Listing~\ref{lst:sample} shows an example output on the 27 sensors used for ADL detection.
It shows the atomic proposition associated with the sensor, the sensor type, the trace file, the fastest change rate (min), and the slowest change rate (max), and whether or not it is skipped.
The rates are provided in milliseconds.
Then, we aggregate over all sensors by computing the fastest and slowest.
Sensors are not included in the aggregate computation (i.e., skipped) if no change appears in their entire trace file.
In this case, we choose 1 second as our polling rate, as no sensor will change twice within a second.

\begin{listing}[t]
	\caption{Rates of change for sensor data. The highlighted sensors are skipped since their data never change.}
	\label{lst:sample}
\begin{minted}{text}
livingroom_table             SensorBool      28.csv     Min: 3000       Max: 230704000   (ms) [OK]
kitchen_dishwasher           SensorThresh    167.csv    Min: 2190810000 Max: 2190810000  (ms) [SKIP]
office_deskplug              SensorThresh    119.csv    Min: 6000       Max: 231159000   (ms) [OK]
office_tv                    SensorBool      283.csv    Min: 420000     Max: 343980000   (ms) [OK]
livingroom_couch             SensorBool      45.csv     Min: 3000       Max: 247031000   (ms) [OK]
kitchen_presence             SensorBool      269.csv    Min: 2000       Max: 230702000   (ms) [OK]
kitchen_c1                   SensorBool      300.csv    Min: 1000       Max: 259080000   (ms) [OK]
kitchen_c2                   SensorBool      315.csv    Min: 1000       Max: 431493000   (ms) [OK]
kitchen_c3                   SensorBool      316.csv    Min: 1000       Max: 259095000   (ms) [OK]
kitchen_c4                   SensorBool      317.csv    Min: 1000       Max: 259051000   (ms) [OK]
kitchen_c5                   SensorBool      355.csv    Min: 1000       Max: 779361000   (ms) [OK]
kitchen_sink_hotwater        SensorThresh    184.csv    Min: 12000      Max: 260085000   (ms) [OK]
kitchen_sink_coldwater       SensorThresh    189.csv    Min: 12000      Max: 260501000   (ms) [OK]
bedroom_closet_door          SensorBool      339.csv    Min: 7000       Max: 605093000   (ms) [OK]
bedroom_luminosity           SensorThresh    120.csv    Min: 1000       Max: 254250000   (ms) [OK]
kitchen_cooktop              SensorThresh    36.csv     Min: 7000       Max: 260333000   (ms) [OK]
bathroom_shower_coldwater    SensorThresh    22.csv     Min: 12000      Max: 345139000   (ms) [OK]
bathroom_shower_hotwater     SensorThresh    201.csv    Min: 12000      Max: 345066000   (ms) [OK]
kitchen_fridge_door          SensorBool      314.csv    Min: 1000       Max: 260749000   (ms) [OK]
livingroom_tv                SensorBool      282.csv    Min: 840000     Max: 344040000   (ms) [OK]
toilet                       SensorThresh    254.csv    Min: 12000      Max: 518222000   (ms) [OK]
bathroom_sink_coldwater      SensorThresh    86.csv     Min: 12000      Max: 260437000   (ms) [OK]
bathroom_sink_hotwater       SensorThresh    264.csv    Min: 25000      Max: 25000       (ms) [SKIP]
kitchen_oven                 SensorThresh    232.csv    Min: 2191235000 Max: 2191235000  (ms) [SKIP]
bedroom_drawer_1             SensorBool      357.csv    Min: 1000       Max: 345825000   (ms) [OK]
bedroom_drawer_2             SensorBool      358.csv    Min: 2000       Max: 515617000   (ms) [OK]
bedroom_bed_pressure         SensorThresh    349.csv    Min: 1000       Max: 342361000   (ms) [OK]

                              (Detected Rate)           Min: 1000       Max: 779361000   (ms)
\end{minted}
\end{listing}

%% file: large.tex
\section{Consideration for Large Traces}
\label{sec:large}
%
Managing the trace length (36,000) is an issue for the monitoring techniques presented in~\cite{themis_issta}.
Since the associated monitors rely on eventual consistency~\cite{CRDT}, in some cases, they wait for input for the length of the trace, which requires a lot of memory.
This was not an issue for the small traces (of length 100) used to compare algorithms originally, but becomes a significantly larger issue when monitoring a real apartment.

Two data structures are introduced in~\cite{themis_issta} to support monitoring decentralized specifications: \emph{memory} and \emph{execution history encoding} ($\MemRep{}$).
We briefly review them in \rsec{sec:datastructures} along with their key operations so we can we present a garbage collection strategy for the memory data structure in \rsec{sec:themis-changes} and an expansion strategy for the $\MemRep{}$ in \rsec{sec:comm-trigger}.
The memory footprint for monitors consists of the sizes of their \emph{memory} and \emph{$\MemRep{}$}.
Both our improvements aim at reducing the size of the data structures for long traces.
Theoretical details for the data structures and monitoring are in~\cite{themis_issta}.
%
\subsection{Monitoring Data Structures and Their Operations}
\label{sec:datastructures}
%
The data structures \emph{memory} and \MemRep{} operate over \emph{atoms}, where an atom is an encoding of atomic propositions.
The encoding used for monitoring the appartment consists of a pair of timestamp and atomic proposition.
For example, the atom $\tuple{23, \mrm{s}_1}$, is used to refer to the truth value of switch $1$ at timestamp $23$.
\paragraph{Memory.}
The \emph{memory} buffers all observations the monitor received from the component it is associated with, and the monitors it depends on.
The memory is a partial function (noted $\mem$) that associates atoms with verdicts.
For example, the memory $\mem = [\tuple{23,\mrm{s}_1} \mapsto \vt, \tuple{23,\mrm{s}_2} \mapsto \vf]$ states that at timestamp 23, switch 1 was enabled while switch 2 was disabled.
An underlying operation used to perform monitoring is denoted by $\seval$, which takes a Boolean expression of atoms, and a memory.
Function $\seval$ attempts to rewrite the expression by replacing the value of the atoms present in the memory by their associated verdict, then simplifies the expression (using Boolean simplification).
The memory stores all observations and is used to rewrite expressions when performing monitoring.
\begin{example}[$\seval$]
	For the expression $e = \tuple{23,\mrm{s}_1} \lor \tuple{23,\mrm{\ell}_1}$ and memory  $\mem = [\tuple{23,\mrm{s}_1} \mapsto \vt]$, applying $\seval(e, \mem)$ will first rewrite $e$ to $\vt \lor \tuple{23,\mrm{\ell}_1}$, which is then simplified to $\vt$.
\end{example}
\paragraph{Execution History Encoding.}
We recall from \rsec{sec:monitors} that monitors are Moore automata that check decentralized traces.
Since we are dealing with partial information due to the decentralized nature of monitors, the $\MemRep{}$ encodes the execution of the underlying automaton, keeping track of potential states when receiving partial observations.
In brief, an $\MemRep{}$ can be modeled as a partial function ($\sys$) that associates a timestamp $t$ and a state $q$ of the automaton with a boolean expression $e$.
Whenever $e$ holds  (i.e., $\sys(t,e)$), we are sure that the automaton is in state $q$ at timestamp $t$.
The Boolean expression $e$ is evaluated using the content of the monitor's \emph{memory} data structure using $\seval$.
The size of the $\MemRep{}$ grows to account for timestamps and potential reachable states as the system executes (as time passes).
The main function that extends the $\MemRep{}$ to new timestamps is $\smove$.
Function $\smove$ takes the current $\MemRep{}$, along with its last stored timestamp, and an arbitrary timestamp in the future, and expands the entries by generating the expressions up to the future timestamp using the structure of the automaton and reachability.
As such, to create an $\MemRep{}$ $\sys'$ from another one $\sys$ containing current information at timestamp $\mathrm{t_{cur}}$ with information up to timestamp $\mathrm{t_{future}}$, we use $\sys' = \smove(\sys, \mathrm{t_{cur}}, \mathrm{t_{future}})$.
Expanding the $\MemRep{}$ when information is missing leads to large expressions in the $\MemRep{}$ which require a larger memory to store and a longer time to simplify.
As such, it is important to ensure that $\smove$ is called when sufficient information is present to resolve the $\MemRep{}$.
%
\subsection{Memory Garbage Collection For Large Traces}
\label{sec:themis-changes}
%
We optimized data structure \emph{memory} (which is used to store observations) to add garbage collection.
To do so we have created a new implementation (\emph{MemoryIndexed}) that indexes observations by timestamp.
When the monitor concludes with a final verdict for timestamp $t$, and respawns to monitor timestamp $t+1$, all observations associated with a timestamp lesser than or equal to $t$ are removed from the memory.
That is, the new memory $\mem'$ is constrained to $\fdom(\mem') = \fdom(\mem) \setminus \setof{\tuple{\mathrm{t'', ap}} \in \fdom(\mem) \mid t'' < t}$ (where $\fdom$ indicates the domain of the partial function).
This ensures that older information is discarded as the monitoring moves with time.
%
\subsection{Lazy $\MemRep{}$ Expansion}
\label{sec:comm-trigger}
%
The $\MemRep{}$ data structure is designed to be as general as possible, and keeps expanding while it has not detected the state the automaton is in.
For large trace sizes, this can cause an $\MemRep{}$ to grow quickly to consume all available memory and prevents monitoring from completion.
That is, the monitor expands the $\MemRep{}$ using $\smove$, causing the expressions to grow exponentially~\cite{tosem}, when no information is provided to the monitor.

This is prominently the case when monitoring safety properties.
Safety properties such as $\mtt{p} \defas \ltlg(\mtt{ap})$ will only conclude when the value of $\mtt{ap}$ is $\vf$.
So long as the value of $\mtt{ap}$ is $\vt$, the monitor checking $\mtt{p}$ does not reach a final verdict, and does not report it to its parent.
Consequently, a monitor that checks a safety property that is never violated, incurs a delay that is as long as the trace size.
%
%
One approach is to limit the expansion of the $\MemRep{}$ to a fixed length (assuming a fixed maximal delay), and use a sliding window to maintain the limit.
This approach, however, may cause monitoring not to conclude in cases where monitoring requires more time than that of the window.
To solve this issue and provide the user with more control, we allow the user to specify the expansion condition for the $\MemRep{}$ as an additional Boolean formula that is determined by communication.
This allows us to expanding the $\MemRep{}$ based on the communication patterns between monitors.
\paragraph{Scope.}
We recall from \rsec{sec:decent} that, for a given monitor labeled $\mathrm{lbl}$, its alphabet $\AP_\mathrm{lbl}$ consists of atomic propositions of dependent monitors and the alphabet of the attached component.
For this enhancement, we consider monitors which only depend on other monitors, i.e., when $\AP_\mathrm{lbl} \subseteq \APmons$.
We can see, when looking at dependencies in \figref{fig:dependencyDAG}, that most monitors eventually rely only on lower-level monitors which themselves rely on component observations.
As such, most high-level specifications for the smart home, and in particular safety properties (formulated as meta-specifications in \rtbl{tbl:props-list}), rely on other monitors which evaluate different specifications, and thus only depend on monitors.
\paragraph{Communication AP.}
For a monitor that only depends on other monitors, its alphabet  consists of monitor references (i.e.,  $\AP_\mathrm{lbl} \subseteq \APmons$).
For each dependent monitor (labeled $\mathrm{dep}$), we create two atomic propositions, one if the received verdict is $\vt$ (noted $\vt_\mathrm{dep}$) and one if it is $\vf$ (noted $\vf_\mathrm{dep}$).
The resulting alphabet is $\AP^\mathrm{com}_\mathrm{lbl} = \setof{\vt_\mathrm{dep}, \vf_\mathrm{dep} \mid dep \in \AP_\mathrm{lbl}}$.
The expansion condition (noted $\varphi^{\mathrm{trigger}}_{\mathrm{lbl}}$) is thus a Boolean expression over the alphabet $\AP^\mathrm{com}_\mathrm{lbl}$.
\paragraph{Evaluating the expansion condition.}
To evaluate the added atomic propositions, we define function $\sadv$ which takes as input an expansion condition $\varphi^{\mathrm{trigger}}_\mathrm{lbl}$, a memory $\mem$, and a timestamp $t$ as follows:
\begin{align*}
	\sadv(\varphi^{\mathrm{trigger}}_\mathrm{lbl}, \mem, t) & = \texttt{ match $\varphi^{\mathrm{trigger}}_\mathrm{lbl}$ with } \\
	                                                        & \begin{array}{ll}
		\mid \top_\mathrm{dep} \in \AP^\mathrm{com}_\mathrm{lbl} & \rightarrow \seval(\tuple{t, \mathrm{dep}}, \mem) = \vt \\
		\mid \bot_\mathrm{dep} \in \AP^\mathrm{com}_\mathrm{lbl} & \rightarrow \seval(\tuple{t, \mathrm{dep}}, \mem) = \vf
	\end{array}
\end{align*}
Function $\sadv$ performs pattern matching to convert the communication atomic proposition to an expression capable of being evaluated using $\seval$, checking if the monitor returned verdict $\vt$ and $\vf$ at timestamp $t$ for $\top_\mathrm{dep}$ and $\bot_\mathrm{dep}$, respectively.
We note that when the atom is not found in the memory, both  $\top_\mathrm{dep}$ and $\bot_\mathrm{dep}$ do not hold.
\paragraph{Triggering the expansion.}
	Given a current time $t_\mathrm{cur}$ for which we last expanded the $\MemRep{}$, we determine the maximum possible expansion for the $\MemRep{}$ by looking for the atom in the memory with the highest timestamp, noted $t_\mathrm{max}$.
Next, we define function $\sadvs$, which takes as input an expansion condition, a memory, a current timestamp and a maximum timestamp and generates the timestamps for which the $\MemRep{}$ must be expanded.
\[
	\sadvs(\varphi^{\mathrm{trigger}}, \mem, t_\mathrm{cur}, t_\mathrm{max}) = \setof{t_\mathrm{cur} < t \leq t_\mathrm{max} \mid \sadv(\varphi^{\mathrm{trigger}}, \mem, t) = \vt}
\]
Finally, we pick the maximum of the timestamps and expand the $\MemRep{}$ accordingly.
\begin{remark}[Wildcard Trigger.]
	\label{rmk:wildcard}
	It is common to observe a expansion condition that involves, for a given monitor (labeled $\mathrm{lbl}$), all the atoms found in the checked specification.
	The expansion condition is then a disjunction of all atoms (i.e., $\bigvee_{\mathrm{ap} \in \AP^\mathrm{com}_\mathrm{lbl}} (\mathrm{ap})$).
	To avoid evaluating such large expression, particularly when many dependencies exist (for example,  meta-specifications $\mathtt{actfloor(0)}$ and $\mathtt{actfloor(1)}$), we provide an optimization flag for a monitor to only trigger expansion upon receiving messages from other monitors.
\end{remark}
\begin{example}[Combination of safety properties and expansion]
	\label{ex:trigger}
	Consider the three monitors $\MONID_0, \MONID_1$ and $\MONID_2$ that check for the following specifications: $\ltlg(\neg \mathtt{firehazard})$, $\ltlg(\neg \mathtt{notwopeople})$ and $\MONID_0 \land \MONID_1$.
	We can see that in this case $\MONID_0$ and $\MONID_1$ only output verdicts when the property is falsified.
	That is, monitor $\MONID_2$ which depends on both, has to normally expand its own $\MemRep{}$ as time passes awaiting information that will only become available when the specification of either is falsified (i.e., $\mathtt{firehazard}$ or $\mathtt{notwopeople}$ evaluate to true in either monitors $\MONID_0$ or $\MONID_1$, respectively).
	As such, we can specify the expansion condition for monitor $\MONID_2$ to be $\vf_{\MONID_0} \lor \vf_{\MONID_1}$: so long as no $\vf$ is communicated from either $\MONID_0$ or $\MONID_1$, the $\MemRep{}$ is not expanded, as it cannot be falsified.
	When using lazy $\MemRep{}$ expansion using  $\MONID_0, \MONID_1$ and $\MONID_2$ with the unbounded operator $\ltlg$, we are able to reduce the maximal observed size of an $\MemRep{}$ during the simulation by as much as 58\%.
\end{example}

%% file: evaluation.tex
\section{Assessing the Monitoring of the Appartment}
%
Monitoring the smart apartment requires leveraging the interdependencies between specifications to be able to scale, beyond monitoring system properties, to more complex meta-specifications (as detailed in \secref{sec:propsgroups}).
We assess using decentralized specifications to monitor the apartment by conducting three scenarios.
The first scenario (\secref{sec:decent-eval}) evaluates the scalability and re-use advantages of using decentralized specifications presented in \secref{sec:decent-advantages} by looking at the complexity of monitor synthesis, and communication and computation costs when adding more complex specifications that re-use sub-specifications.
The second scenario (\secref{sec:decent-limitations}) evaluates the effectiveness of detecting ADL by looking at various detection measures such as precision and recall.
The third scenario (\secref{sec:adaptation}) portrays the advantages of modularity by (i) adapting specification $\mtt{napping}$ to use different sensors without modifying dependencies, and (ii) porting specification $\mtt{firehazard}$ to a completely different environment (using the ARAS dataset~\cite{ARAS}).

%
\subsection{Monitoring Efficiency and Hierarchies}
\label{sec:decent-eval}
%
\paragraph{Monitor synthesis.}
Table~\ref{tbl:props-list} displays the number of atomic propositions referenced by each specification for the decentralized ($|\mathrm{AP}^\mathrm{d}|$) and the centralized ($|\mathrm{AP}^\mathrm{c}|$) settings.
Column $\mathrm{d}$ indicates the maximum depth of the directed acyclic graph of dependencies.
We use the depth to assess how many levels of sub-specifications need to be computed.
When $d=0$, it indicates that the specification can be evaluated directly by the monitor placed on the component, while $d=1$ indicates that the monitor has to poll at most 1 monitor for its verdict (which typically relays the component observations).
More generally, when $d=n$, it indicates that the specification depends on a monitor that has at most depth $n-1$.
The atomic propositions indicate either direct references to sensor observations (in the centralized setting) or references to either sensor observations or dependent monitors (in the decentralized setting).
For certain specifications such as $\mtt{toilet}$ which relies only on the water sensor in the toilet to be detected, there is no difference between using a centralized or decentralized specification, as it resolves to the observations.
Reduction becomes more pronounced when specifications re-use other specifications as sub-specifications.
For example, specification $\mtt{acthouse} \defas \mtt{actfloor}(0) \lor \mtt{actfloor}(1)$, when decentralized, uses only 2 references (for each of the sub-specifications).
However, when expanded, it references all 27 sensors used to detect activities.
Additionally, specification $\mtt{notwopeople} \defas \neg(\mtt{actfloor}(0) \land \mtt{actfloor}(1))$ would not re-use the sub-specifications if expanded, requiring all sensors again.
Henceforth, re-use greatly reduces the formula size and allows us to synthesize the monitors needed to check the formulas, as the synthesis algorithm is doubly exponential as mentioned in \secref{sec:decent-advantages}.

\begin{figure}[t]
  \centering
  \subfloat[Communication.]{
       \centering
       \includegraphics[width=.5\linewidth]{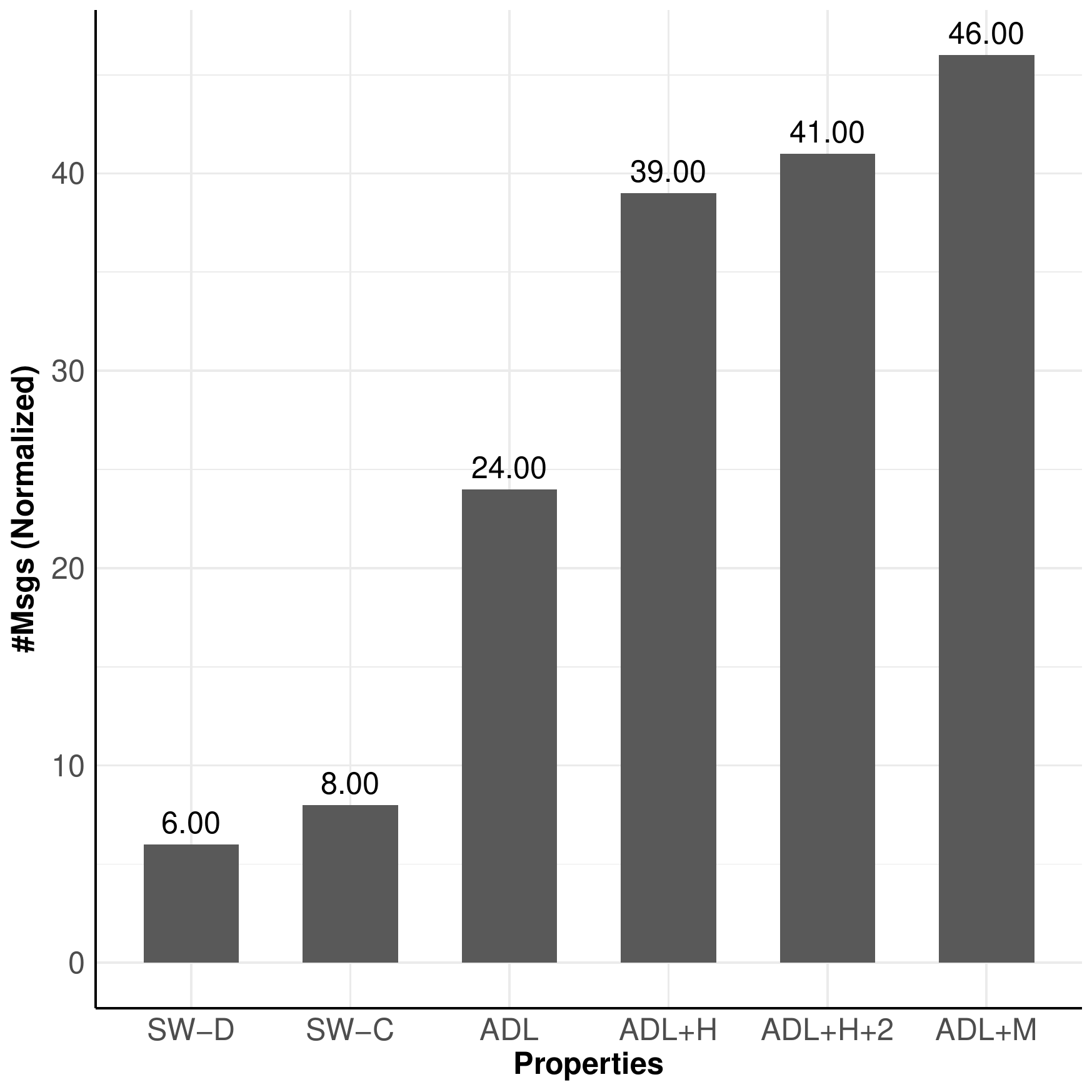}
       \label{fig:perf-msgnum}
  }
  \subfloat[Computation.]{
       \centering
       \includegraphics[width=.5\linewidth]{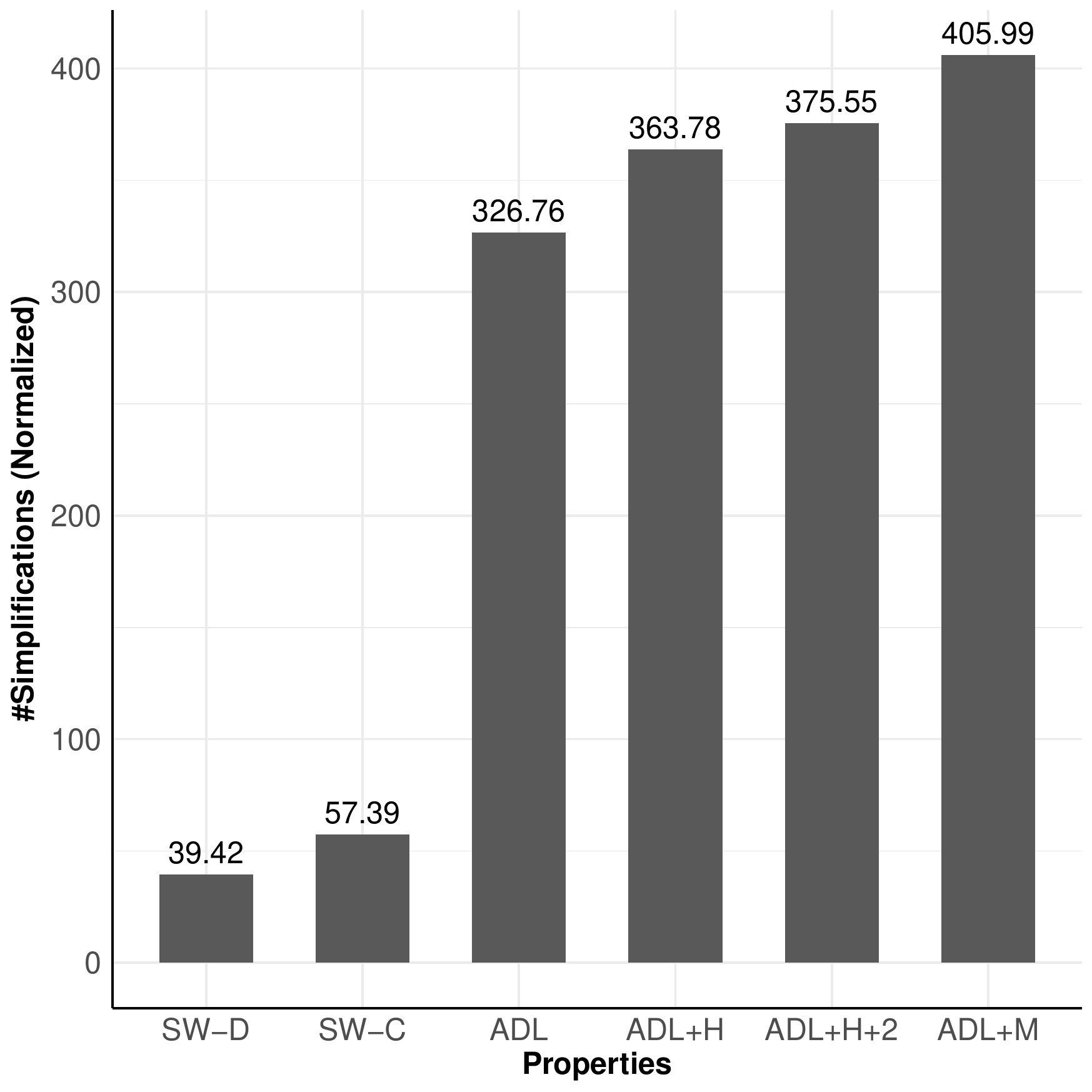}
       \label{fig:perf-simp}
  }
  \caption{%
  Scalability of communication and computations in decentralized specifications.}
  \label{fig:perf}
\end{figure}
\paragraph{Assessing re-use and scalability.}
Reducing the size of the atomic propositions needed for a specification not only affects monitor synthesis, but also runtime performance, as atomic propositions represent the information needed to determine the specification (\secref{sec:decent-advantages}).
To assess re-use and scalability, we perform two tasks and gather two measures pertaining to computation and communication, and present results in \figref{fig:perf}.
The first task compares a centralized (SW-C) and a decentralized (SW-D) version of specification  $\mtt{sc\_ok}$ presented in Example~\ref{ex:sclight-synthesis} using only 2 rooms.
The second task introduces large meta-specifications on top of the ADL specifications to check scalability.
Firstly, we measure the communication and computation for monitoring ADL specifications (ADL).
Secondly, we introduce specifications $\mtt{actfloor}(0)$, $\mtt{actfloor}(1)$ and $\mtt{acthouse}$ (ADL+H) as they require information about all sensors for ADL.
Thirdly, we add specification $\mtt{notwopeople}$ (ADL+H+2), as it re-uses the same sub-specifications as  specification $\mtt{acthouse}$.
Lastly, we show all measures for all meta-specifications in Table~\ref{tbl:props-list} (ADL+M).
We re-use two measures from~\cite{themis_issta}: the total number of simplifications the monitors are doing, and the total number of messages transferred.
These measures are provided directly with \texttt{THEMIS}~\cite{themis_tool}.
The total number of messages abstracts the communication (\textbf{\#Msgs}), as our messages are of fixed length, they also represent the total data transferred.
The total number of simplifications (\textbf{\#Simplifications}) abstracts the computation done by the monitors, as they attempt to simplify Boolean expressions that represent automaton states, which are the basic operations for maintaining the monitoring data structures in~\cite{themis_issta}.
Both measures are normalized by the number of timestamps in the execution (36,000).
The resulting normalized measures represent the number of simplifications and messages per round.
\paragraph{Results.}
Figure~\ref{fig:perf-msgnum} shows the normalized number of messages sent by all monitors.
For the first task, we notice that the number of messages is indeed lower in the decentralized setting, SW-D sends on average 2 messages per timestamp less than SW-C, which corresponds to the difference in the number of atomic propositions referenced (6 for SW-D and 8 for SW-C).
For the second task, we notice that on the baseline for ADL, we observe 24 messages per timestamp, a smaller number than the sensors count (27).
This is because some ADL like $\mtt{toilet}$ are directly evaluated on the sensor without communicating, and other ADL like $\mtt{preparing}$, re-use other ADL specifications like $\mtt{kactivity}$.
By introducing the 3 meta-specifications stating that an activity occurred on a floor or globally in the appartment, the number of messages per round only increases by 15.
This also coincides with the number of atomic propositions for the specifications (6 for $\mtt{actfloor}(0)$, 7 for $\mtt{actfloor}(1)$, and 2 for $\mtt{acthouse}$) as those monitors depend in total on 15 other monitors to relay their verdicts.
This costs much less than polling 16 sensors to determine $\mtt{actfloor}(0)$, 11 sensors to determine $\mtt{actfloor}(1)$, and 27  (a total of 54) to determine $\mtt{acthouse}$.
To verify this, we notice that the addition of $\mtt{notwopeople}$ (ADL+H+2) that needs information from all 27 sensors, only increases the total number of messages per timestamp by 2.
The specification $\mtt{notwopeople}$ reuses the verdicts of the two monitors associated with each $\mtt{actfloor}$ specification.
After adding all the meta-specifications (ADL+M), the total number of messages per timestamp is 46, whihc is less than the number needed to verify adding $\mtt{actfloor}$, and $\mtt{acthouse}$ in a centralized setting (54).
We notice a similar effect for computation (\figref{fig:perf-simp}).

%
\subsection{ADL Detection using RV}
\label{sec:decent-limitations}
%
\begin{table}[t]
  \newcommand\prna[0]{\multicolumn{3}{c}{-}}
  \centering
  \caption{Precision, Recall, and F1 scores of monitoring all ADL specifications on three days with different schedules.}
  \label{tbl:props-accuracy-measures}
  \begin{tabular}{l c c c c c c c c c}\toprule
      & \multicolumn{3}{c}{\scriptsize{Tuesday, Jan 31 2017}}
      & \multicolumn{3}{c}{\scriptsize{Monday, Feb 20 2017}}
      & \multicolumn{3}{c}{\scriptsize{Tuesday, Feb 21 2017}}
      \\
      \textbf{Specification}
      & \textbf{Precision} & \textbf{Recall} & \textbf{F1}
      & \textbf{Precision} & \textbf{Recall} & \textbf{F1}
      & \textbf{Precision} & \textbf{Recall} & \textbf{F1}
      \\

      \midrule
      $\mtt{computing}$
        &  0.98  & 0.99 & 0.99
        &  0.94  & 0.99 & 0.96
        &  0.99  & 0.99 & 0.99
      \\
      $\mtt{office\_tv}$
        & 1.00 & 0.80 & 0.89
        & 1.00 & 0.94 & 0.97
        & \prna
      \\
      $\mtt{cooking}$
        &  0.88 & 0.88 & 0.88
        &  0.90 & 0.93 & 0.92
        &  \prna
      \\
      \midrule
      $\mtt{shower\_usage}$
      &  1.00 & 0.50 & 0.67
      &  \prna
      &  1.00 & 0.63 & 0.77
      \\
      $\mtt{washing\_dishes}$
      & 1.00 & 0.47 & 0.64
      & 0.93 & 0.63 & 0.75
      & \prna
      \\
      $\mtt{livingroom\_tv}$
      & 1.00 & 0.43 & 0.60
      & \prna
      & 1.00 & 0.47 & 0.64
      \\
      $\mtt{dressing}$
      & 1.00 & 0.41 & 0.58
      & 1.00 & 0.31 & 0.47
      & \prna
      \\
      \midrule
      $\mtt{toilet}^*$
      &  1.00 & 0.18 & 0.30
      &  \prna
      &  0.75 & 0.24 & 0.36
      \\
      $\mtt{sink\_usage}$
      &  1.00 & 0.13 & 0.23
      &  1.00 & 0.24 & 0.35
      &  0.003 & 0.16 & 0.01
      \\
      \midrule
      $\mtt{eating}$
      & 0.61 & 0.35 & 0.44
      & 0.70 & 0.73 & 0.71
      & \prna
      \\
      \midrule
      $\mtt{napping}$
      &  0.43 & 0.95 & 0.60
      &  0.38 & 0.94 & 0.54
      &  \prna
      \\
      $\mtt{preparing}$
      & 0.23 & 0.77 & 0.35
      & 0.21 & 0.79 & 0.34
      & \prna
      \\
      \midrule
      $\mtt{reading}$
      &  0.37 & 0.04 & 0.06
      &  0.02 & 0.10 & 0.03
      &  \prna
      \\

       \bottomrule
  \end{tabular}
\end{table}

\paragraph{Measurements.}
Table~\ref{tbl:props-accuracy-measures} displays the effectiveness of using RV to detect all ADL specifications on the trace of three days with different schedules.
To assess the effectiveness, we compared with the provided self-annotated data from~\cite{orangehome}, where the user annotated the start and end of each activity.
We measure precision, recall and F1 (the geometric mean of precision and recall).
To measure precision, we consider a true positive when the verdict $\vt$ of a monitor for a given timestamp fell indeed in the self-annotated interval for the activity.
To measure recall, we measure the proportion of the intervals for which the monitors have determined $\vt$ (using RV).
This approach is more fine-grained than the approach used in~\cite{Paula} where the precision and recall are computed for the start and end of intervals.
\paragraph{Results.}
The effectiveness of detection depends highly on the specification.
Our approach performs well for the specifications $\mtt{computing}$, $\mtt{cooking}$, $\mtt{office\_tv}$, as it exhibits high precision and high recall.
The second group of specifications contains specifications such as $\mtt{shower\_usage}$, and $\mtt{livingroom\_tv}$.
It exhibits high precision but medium recall, that is, we were able to determine around 40 to 50\% of all the timestamps where the specifications held according to the person annotating, without any false positives.
The third group is similar to the second group but has very low recall (13-18\%) and contains the specifications $\mtt{toilet}$ and $\mtt{sink\_usage}$.
We notice that for $\mtt{sink\_usage}$ specific user behavior can throw it off, as seen for the trace of Feb 21, we elaborate on the limitations in the next paragraph.
The fourth group, which includes the specifications $\mtt{napping}$ and $\mtt{preparing}$, shows high recall but a high rate of false positives.
And finally, specification $\mtt{reading}$ is not properly detected, as it has a high rate of false positives and covers almost no annotated intervals.
\paragraph{Limitations of RV for detecting ADL.}
The limitations of using RV to detect ADL are due to the modeling.
As mentioned in \secref{sec:propsgroups}, RV can be seen as a knowledge-based approach to activity detection, as such it suffers from similar weaknesses and limitations~\cite{chen_adl}.
The activity is described as a rigid formal specification over the sensor data, and this has two consequences.
Firstly, since RV relies purely on sensor data, activities which cannot be inferred from existing sensors will be poorly detected or not detected at all.
This is the case for $\mtt{reading}$, as there are no sensors to indicate that the tenant is reading.
We infer reading by checking that the light is on in the room and no other specified activity holds.
Secondly, given that specifications are rigid, we expect the user to behave exactly as specified for the activity to be detected, any minor deviation results in the activity not being detected (as seen on Feb 21).
To illustrate this point, the specification $\mtt{computing}$ relies on the power consumption of the plug in the office.
Had the tenant been charging his phone instead of computing, the recall would have suffered greatly.
Another great example of this is the $\mtt{shower\_usage}$ specification, that is captured by inspecting the water usage of the shower.
The time the tenant spends getting into the shower and out of the shower will not be considered, which greatly impacts recall.
The above issues are further compounded by the annotation being carried out by a person.
The annotator can for example take a few seconds to annotate some events which could impact recall, especially for short intervals of activity.
However, even with the inherent limitations of using knowledge-based approaches, our observed groups and results fall within the expected range, of knowledge-based approaches such as~\cite{Paula}, and also have similar effectiveness as model-based SVM approaches such as~\cite{chen_svm}.
We elaborate on how the introduced modularity from decentralized specifications can alleviate some of these issues in \secref{sec:adaptation}.
%
\subsection{Specification Adaptation for ADL Detection}
\label{sec:adaptation}
%
Decentralized specifications introduce numerous advantages (see \secref{sec:decent-advantages}) for monitoring hierarchical systems that can change.
We illustrated in \secref{sec:decent-eval} the scalability of decentralized specifications with hierarchies.
Decentralized specifications allows specifications to be written with references to other specifications.
The references allow specifications to be modular, changing the referenced specification is transparent with no modification to the specifications that depend on it.
In this section, we illustrate the advantages of modularity in two cases.
In the first case, we improve the detection of the activity $\mtt{napping}$ by adding relevant sensors.
The change only requires changing the monitor for $\mtt{napping}$, and no change is necessary for the remaining dependent specifications.
In the second case, we apply the specification $\mtt{firehazard}$ and all its dependencies on a completely different environment using the ARAS dataset~\cite{ARAS}.
\begin{table}[t]
  \centering
  \captionsetup{position=top}
  \caption{Modifying the decentralized specification to improve detection, and adapt to new environment.}
  \label{tbl:modularity}
  \subfloat[Refining $\mtt{napping}$ using the bedroom sensors:  bed pressure ($\mtt{weight}$), presence ($\mtt{pres}$), and light  ($\mtt{\ell}$).]{%
      \label{tbl:props-napping}%
      \parbox{0.5\textwidth}{\begin{tabular}{l c c c }\toprule

          \textbf{Formula}
          & \textbf{Precision} & \textbf{Recall} & \textbf{F1}
          \\

          \midrule
          $\ltlgw{25}(\mtt{weight})$ &  0.43 & 0.95 & 0.60\\
          $\ltlgw{3}(\mtt{weight})$ & 0.43 & 0.99 & 0.60\\
          $\ltlfw{3}(\mtt{weight})$ &  0.43 & 1.0 & 0.60\\
          \midrule
          $\ltlgw{3}(\mtt{pres} \land \mtt{weight})$ &  0.34 & 0.14 & 0.20\\
          $\ltlgw{3}(\neg\mtt{\ell} \land \mtt{weight})$ &  1.00 & 0.97 & 0.99\\
           \bottomrule
      \end{tabular}}%
  }\hfill%
  \subfloat[Modifications to detect $\mtt{firehazard}$ in ARAS.]{%
  \label{tbl:props-changes}%
    \parbox{0.5\textwidth}{\begin{tabular}{ll} \toprule
    \textbf{Specification} & \textbf{Formula}\\
    \midrule
    $\mtt{preparing}$ & $\ltlfw{3}(\mtt{m\_kdrawer} \lor \mtt{m\_fridge} \lor \mtt{m\_cupboard})$\\
    $\mtt{cooking}$ & $\mtt{preparing}$\\
    \midrule
    $\mtt{beds}$    & $\mtt{bed1} \lor \mtt{bed2}$\\
    $\mtt{beds}'$   & $\mtt{bed1} \land \mtt{bed2}$\\
    $\mtt{napping}$ & $\ltlgw{25}(\mtt{beds})$\\
    \midrule
    $\mtt{firehazard}$ & $\mtt{napping} \implies  \neg\mtt{cooking}$\\
    \bottomrule
  \end{tabular}}%
  }
\end{table}

\paragraph{Improving activity detection.}
We modify the specification $\mtt{napping}$ to better capture the activity.
This requires no change to specifications that depend on $\mtt{napping}$.
Table~\ref{tbl:props-napping} shows the changes in precision and recall, for various versions of the specification $\mtt{napping}$.
We modify the formula to relax the time constraints on the output of the bed pressure sensor.
We notice, that while this could slightly improve recall (0.95 to 1), it does not translate to any precision improvement (it remains at 0.43).
We explore using additional sensors in the room to capture the specification better.
Using the presence sensor proves to be detrimental as it reduces precision to 0.34 and recall to 0.14.
This is reasonable, as the presence sensor is a motion detector, and when someone is sleeping there may be no motion at all.
However, people typically tend to turn the lights off when sleeping.
Using the additional light sensor to detect lights are off, helps us increase precision to 1 and recall to 0.99.
One could see that the effect of ADL detection is behavior specific, a tenant that sleeps with lights on will have undetected sleep using our specification.
Being able to change to specific parts of the specification without impacting the rest of the it provides the flexibility to tune the ADL detection to specific users and behaviors.
\paragraph{Adapting to new environments.}
In \secref{sec:propsgroups} we mentioned that ADL can be challenging as the detection of the specification does not only depend on the user behavior, but also on the environment in which it is monitored.
In the context of learning techniques, using information learned from one environment to apply it to detection of ADL in other environments is discussed in~\cite{KasterenEK10}.
Since decentralized specifications provide both a hierarchical and modular approach to designing specifications, it is possible to adapt specifications to new environment, by only changing the relevant parts or dependencies, and reasoning at the appropriate level.
For instance, while specifications specifying ADL may change depending on the sensors and user behavior, meta-specifications do not necessarily change.
We adapt specification $\mtt{firehazard}$ and all its dependencies in the ARAS~\cite{ARAS} dataset.
The ARAS dataset features contact, pressure, distance, and light sensors, recording the interactions of two tenants with the sensors over a period of 30 days.

Table~\ref{tbl:props-changes} shows the changes in the decentralized specification compared with that of Amiqual4Home found in Appendix~\ref{appendix:properties}.
For activity $\mtt{preparing}$, we follow a similar pattern, looking at the usage of cupboards, fridge, and kitchen drawers.
Thus, we adapt the formula to reflect the available sensors in the kitchen.
However, the ARAS dataset does not provide any electricity sensors for appliances, nor any way to detect heat being turned on.
As such it is impossible to detect cooking using any sensors.
Since we cannot tell $\mtt{preparing}$ and $\mtt{cooking}$ apart, we define $\mtt{cooking}$ to simply be equivalent to $\mtt{preparing}$.
Notice how in this case, we inverted the dependency from \figref{fig:dependencyDAG} (in ARAS, $\mtt{cooking}$ depends on $\mtt{preparing}$).
The ARAS dataset records the behavior of \emph{two} people, instead of just \emph{one}.
As such, activity $\mtt{napping}$ needs to be adjusted for the \emph{two beds}.
There are two ways to do so, the first assumes either one of the tenants is napping ($\mtt{beds}$), and the second assumes both are napping simultaneously ($\mtt{beds}'$).
We notice that the meta-specification $\mtt{firehazard}$ remains unchanged.
However, it has two different interpretations.
If we use $\mtt{beds}$, then it is possible to trigger $\mtt{firehazard}$ when one tenant is cooking while the other is sleeping.
We verify that, and notice that it is indeed falsified in 8 days (7, 9, 16, 17-19, 24, 27).
Using $\mtt{beds}'$, allows us to only capture $\mtt{firehazard}$ when both tenants are sleeping.
It is then possible to refer $\mtt{napping}$ to $\mtt{allnapping}$ and $\mtt{anynapping}$, then using $\mtt{firehazard}$ on $\mtt{allnapping}$, which would apply in both scenarios.
\paragraph{Discussion.}
We see that modularity provides several advantages.
It allows us to make local change to specifications that do not need to be propagated upwards.
It also makes it possible to generalize and abstract the specification to adapt to multiple environments.
Decentralized specifications allow specifications to be written in a modular and adaptable fashion, allowing specifications to be adapted to target changes in user behavior and environment.
It can be seen much like component-based design~\cite{CBSE}, which separate the implementation of each component in software, from its interaction  with other components.

%% file: related.tex
\section{Related Work}
%
We present similar or useful techniques for detecting ADL in a smart apartment that use log analysis and complex event processing.
Then, we present techniques from stream-based RV that can be extended for monitoring smart apartments.
\paragraph{ADL detection using log analysis.}
Detecting ADL can be performed using trace analysis tools.
The approach in~\cite{Paula} defines parametric events using Model Checking Language (MCL)~\cite{MCL} based on the modal mu-calculus (inspired by temporal logic and regular expressions).
Traces are read and transformed into actions, then actions are matched against the specifications to determine locations in the trace that match ADL.
Five ADL (sleep, using toilets, cooking, showering, and washing dishes) are specified and checked in the same smart apartment as our work.
While this technique is able to detect ADL activities, it amounts to checking traces offline, and a high level of post-processing is required to analyze the data.
In~\cite{BasinCEHKM16}, the authors describe an approach for log analysis at very large scale.
The specification is expressed using Metric First Order Temporal Logic (MFOTL), and logs are expressed as a temporal structure.
The authors develop a \emph{MapReduce} monitoring algorithm to analyze logs generated by more than 35,000 computers, producing approximately 1 TB of log data each day.
While this approach is designed for distributed systems, does not map dependencies, and works offline, it could be used to process and monitor rich specifications over sensor data seen as log files.
\paragraph{ADL detection using Complex Event Processing.}
Reasoning at a much higher level of abstraction than sensor data, the approach in~\cite{ADLBEEPBEEP} attempts to detect ADL by analyzing the electrical consumption in the household.
To do so, it employs techniques from Complex Event Processing (CEP), in which data is fed as streams and processed using various functions to finally output a stream of data.
In this work, the ADL detection is split into two phases, one which detects peaks and plateaus of the various electrical devices, and the second phase uses those to indicate whether or not an appliance is being used.
This illustrates a transformation from low-level data (sensor signal) to a high-level abstraction (an appliance is being used).
The use of CEP for detecting ADL is promising, as it allows for similar scalability and abstraction.
However, CEP's model of named streams makes it hard to analyze the specification formally, making little distinction between specification and implementation of the monitoring logic.
\paragraph{ADL detection using Runtime Verification.}
Similarly to CEP but focusing on Boolean verdicts, various stream-based RV techniques have been elaborated such as LOLA~\cite{LOLA} which are used to verify correctness properties for synchronous systems such as the PCI bus protocol and a memory controller.
A more recent approach uses the Temporal Stream-Based Specification Language (TeSSLa) to verify embedded systems using FPGAs~\cite{COEMS}.
Stream-based RV is particularly fast and effective for verifying lengthy parametric traces.
However, it is unclear how these approaches handle monitor synthesis for a large number of components and account for the hierarchy in the system.
\paragraph{Discussion.}
Stream-based systems such as stream-based RV and CEP are bottom-up.
Data in streams is eventually aggregated into more complex information and relayed to a higher level.
Decentralized specifications also support top-down approaches, which would increase the efficiency of monitoring large and hierarchical systems.
To illustrate the point, consider the decentralized specification in \figref{fig:sclight}.
In the automaton $\aut_{\mtt{sc\_light}_i}$, the evaluation of the dependent monitor $\aut_{\mrm{\ell}_i}$ only occurs when reaching $q_1$, so long as the automaton is in $q_0$, no interaction with the dependent monitor is necessary.
This top-down feedback can be used to naturally optimize dependencies and increase efficiency.
Because of the oracle-based implementation of decentralized specifications, it is possible to integrate any monitoring reference that eventually returns a verdict.
One could imagine integrating other stream-based monitors or even data-driven ADL detection approaches.
The integration works both ways, as monitors can be considered a (blocking) stream of verdicts for the other techniques.

%% file: conclusion.tex
\section{Conclusion and Future Work}
\label{sec:conclusion}
%
\subsection{Conclusion}
%
Monitoring a smart apartment presents RV with interesting new problems as it requires a scalable approach that is compositional, dynamic, and able to handle a multitude of devices.
This is due to the hierarchical structure imposed by either limited communication capabilities of devices across geographical areas or the dependencies between various specifications.
Attempting to solve such problems with centralized specifications is met with several obstacles at the level of monitor synthesis techniques (as we are presented with large formulae), and also at the level of monitoring as one needs to model interdependencies between formulae and re-use the sub-specifications used to build more complex specifications.
We illustrate how decentralized specifications tackle such systems by explicitly modeling of interdependencies between specifications.
Furthermore, we illustrate monitoring specifications that detect ADL in addition to system properties and even more specifications defined over both types of specifications.
%
\subsection{Future Work}
%
We believe that the use of decentralized specifications could be further extended to bring monitoring closer to data (collected on sensors), and make RV a suitable verification technique for edge computing.
One challenge of the case study was to determine the correct sampling period for monitor to operate.
Further investigation is required to layout the tradeoffs between the sampling period, communication overhead, and energy consumption.
Also, decentralization is only supported by specifications based on the standard (point-based) LTL3 semantics.
We believe that the use and decentralization of richer specification languages are desirable.
For instance, we consider (i) using a counting semantics able to compute the number of steps needed to witness the satisfaction or violation of a specification~\cite{abs-1804-03237} (ii) using techniques allowing to deal with uncertainty (e.g., in case of message loss)~\cite{BartocciG13} (iii) using spatio-temporal specifications (e.g.~\cite{Haghighi:2015}) to reason on physical locations in the house, and (iv) using a quantitative semantics possibly with time~\cite{BakhirkinFMU17}.
Finally, we consider using runtime  enforcement~\cite{Falcone10,FalconeMFR11,FalconeMRS18} techniques (especially those for timed specifications~\cite{FalconeJMP16}) to guarantee system properties and improve safety in the house (e.g., disabling cooking equipment whenever specification $\mtt{firehazard}$ is violated).
This requires to define the foundations for decentralized runtime enforcement on the theoretical side, and provide houses and monitors with actuators on the practical side.

%% file: appendix.tex
\section{List of Properties}
\label{appendix:properties}

Table~\ref{tbl:props-def} shows all property definitions used in this case study.
We ommitted the smaller monitors that are trivial such as $\mtt{m\_kitchen\_cupboard}$ which is a disjunction of all cupboard doors observations in the kitchen.

\input{table-props}

%% file: table-props.tex
\begin{table}[h!]
  \centering
  \caption{%
  Definitions of the specifications used in the case study.
  A specification with name prefixed with $\mtt{m\_}$ is such that the corresponding monitor is directly deployed on the component.
  }
  \label{tbl:props-def}

  \begin{tabular}{l p{9.5cm}}\toprule

      \textbf{Name} & \textbf{Formula}\\
      \midrule

      $\mtt{sc\_light}(i)$ & $\ltlg(\mathrm{switch}_i \implies \ltlx(\mathrm{light}_i \ltlu \neg\mathrm{switch}_i), i \in [0..3]$\\
      $\mtt{sc\_ok}$ & $\bigwedge_{i \in [0 .. 3]} \mtt{sc\_light}(i)$\\
      \midrule

      $\mtt{m\_toilet}$ & $\mtt{toilet\_water}$\\

      $\mtt{sink\_usage}$ & $\ltlgw{3}(\mtt{m\_bathroom\_sink\_water})$ \\
      $\mtt{m\_bathroom\_sink\_water}$ & $\mtt{bathroom\_sink\_cold} \lor \mtt{bathroom\_sink\_hot}$\\

      $\mtt{shower\_usage}$ & $\ltlgw{2}(\mtt{m\_bathroom\_shower\_water})$\\
      $\mtt{napping}$ & $\ltlgw{25}(\mtt{m\_bedroom\_bed\_pressure})$\\
      $\mtt{dressing}$ & $\ltlfw{4}(\mtt{m\_bedroom\_closet\_door} \lor \mtt{m\_bedroom\_drawers}))$\\
      $\mtt{reading}$ & $\mtt{m\_bedroom\_light} \land \ltlfw{4}(\neg\mtt{dressing} \land \neg\mtt{napping})$\\
      $\mtt{office\_tv}$ & $\ltlfw{3}(\mtt{m\_office\_tv})$\\
      $\mtt{computing}$ & $\ltlfw{3}(\mtt{m\_office\_deskplug})$\\
      $\mtt{cooking}$ & $\ltlfw{5}(\mtt{m\_kitchen\_cooktop} \lor \mtt{m\_kitchen\_oven})$\\
      $\mtt{washing\_dishes}$ & $\ltlfw{3}(\mtt{m\_kitchen\_dishwasher} \lor \mtt{m\_kitchen\_sink\_water})$\\
      $\mtt{kactivity}$ & $\mtt{m\_kitchen\_presence} \land \ltlfw{3}(\mtt{m\_kitchen\_sink\_water} \lor \mtt{m\_kitchen\_fridgedoor} \lor \mtt{m\_kitchen\_cupboard})$\\
      $\mtt{preparing}$ & $\mtt{kitchen\_activity} \land \neg\mtt{cooking}$\\
      $\mtt{livingroom\_tv}$ & $\ltlfw{3}(\mtt{m\_livingroom\_tv} \land \mtt{m\_livingroom\_couch})$\\
      $\mtt{eating}$ & $\neg\mtt{m\_kitchen\_presence} \land \ltlgw{6}(\mtt{m\_livingroom\_table})$\\

      \midrule
      $\mtt{actfloor}(0)$ & $\mtt{cooking} \lor \mtt{preparing} \lor \mtt{eating} \lor \mtt{washing\_dishes} \lor \mtt{livingroom\_tv} \lor \mtt{m\_toilet}$\\
      $\mtt{actfloor}(1)$ & $\mtt{computing} \lor \mtt{dressing} \lor \mtt{napping} \lor \mtt{office\_tv} \lor \mtt{reading} \lor \mtt{shower\_usage} \lor  \mtt{sink\_usage}$\\
      $\mtt{acthouse}$ & $\mtt{actfloor}(0) \lor \mtt{actfloor}(1)$\\
      $\mtt{notwopeople}$ & $\neg(\mtt{actfloor}(0) \land \mtt{actfloor}(1))$\\
      $\mtt{restricttv\_office}$ & $\mtt{office\_tv} \implies \ltlfw{10}(\neg\mtt{office\_tv})$\\
      $\mtt{restricttv\_living}$ & $\mtt{livingroom\_tv} \implies \ltlfw{10}(\neg\mtt{livingroom\_tv})$\\
      $\mtt{restricttv}$ & $\mtt{restricttv\_living} \land \mtt{restricttv\_office}$\\
      $\mtt{firehazard}$ & $\mtt{napping} \implies  \neg\mtt{cooking}$\\

       \bottomrule

  \end{tabular}

\end{table}

%% file: main.bbl
\begin{thebibliography}{10}
\providecommand{\url}[1]{\texttt{#1}}
\providecommand{\urlprefix}{URL }

\bibitem{home_energy}
Aimal, S., Parveez, K., Saba, A., Batool, S., Arshad, H., Javaid, N.: Energy
  optimization techniques for demand-side management in smart homes. In:
  Advances in Intelligent Networking and Collaborative Systems, The 9th
  International Conference on Intelligent Networking and Collaborative Systems,
  INCoS-2017. Lecture Notes on Data Engineering and Communications
  Technologies, vol.~8, pp. 515--524. Springer (2017)

\bibitem{ARAS}
Alemdar, H.{\"{O}}., Ertan, H., Incel, {\"{O}}.D., Ersoy, C.: {ARAS} human
  activity datasets in multiple homes with multiple residents. In: 7th
  International Conference on Pervasive Computing Technologies for Healthcare
  and Workshops, PervasiveHealth 2013. pp. 232--235. {IEEE} (2013)

\bibitem{DBLP:conf/issta/2017}
Proceedings of the 26th {ACM} {SIGSOFT} International Symposium on Software
  Testing and Analysis, Santa Barbara, CA, USA, July 10 - 14, 2017. {ACM}
  (2017)

\bibitem{BakhirkinFMU17}
Bakhirkin, A., Ferr{\`{e}}re, T., Maler, O., Ulus, D.: On the quantitative
  semantics of regular expressions over real-valued signals. In: Abate, A.,
  Geeraerts, G. (eds.) Formal Modeling and Analysis of Timed Systems - 15th
  International Conference, {FORMATS} 2017, Berlin, Germany, September 5-7,
  2017, Proceedings. Lecture Notes in Computer Science, vol. 10419, pp.
  189--206. Springer (2017)

\bibitem{abs-1804-03237}
Bartocci, E., Bloem, R., Nickovic, D., R{\"{o}}ck, F.: A counting semantics for
  monitoring {LTL} specifications over finite traces. CoRR  abs/1804.03237
  (2018)

\bibitem{lncs/10457}
Bartocci, E., Falcone, Y. (eds.): Lectures on Runtime Verification -
  Introductory and Advanced Topics, Lecture Notes in Computer Science, vol.
  10457. Springer (2018)

\bibitem{Bartocci2017}
Bartocci, E., Falcone, Y., Bonakdarpour, B., Colombo, C., Decker, N., Havelund,
  K., Joshi, Y., Klaedtke, F., Milewicz, R., Reger, G., Rosu, G., Signoles, J.,
  Thoma, D., Zalinescu, E., Zhang, Y.: First international competition on
  runtime verification: rules, benchmarks, tools, and final results of crv
  2014. International Journal on Software Tools for Technology Transfer  (Apr
  2017)

\bibitem{BartocciFFR18}
Bartocci, E., Falcone, Y., Francalanza, A., Reger, G.: Introduction to runtime
  verification. In: Bartocci, E., Falcone, Y. (eds.) Lectures on Runtime
  Verification - Introductory and Advanced Topics, Lecture Notes in Computer
  Science, vol. 10457, pp. 1--33. Springer (2018),
  \url{https://doi.org/10.1007/978-3-319-75632-5\_1}

\bibitem{BartocciG13}
Bartocci, E., Grosu, R.: Monitoring with uncertainty. In: Bortolussi, L.,
  Bujorianu, M.L., Pola, G. (eds.) Proceedings Third International Workshop on
  Hybrid Autonomous Systems, {HAS} 2013, Rome, Italy, 17th March 2013. {EPTCS},
  vol. 124, pp. 1--4 (2013)

\bibitem{BasinCEHKM16}
Basin, D.A., Caronni, G., Ereth, S., Harvan, M., Klaedtke, F., Mantel, H.:
  Scalable offline monitoring of temporal specifications. Formal Methods in
  System Design  49(1-2),  75--108 (2016)

\bibitem{FAMTL}
Basin, D.A., Klaedtke, F., Zalinescu, E.: Failure-aware runtime verification of
  distributed systems. In: Harsha, P., Ramalingam, G. (eds.) 35th {IARCS}
  Annual Conference on Foundation of Software Technology and Theoretical
  Computer Science, {FSTTCS} 2015. LIPIcs, vol.~45, pp. 590--603. Schloss
  Dagstuhl - Leibniz-Zentrum fuer Informatik (2015)

\bibitem{SALTNFM}
Bauer, A., Leucker, M.: The theory and practice of {SALT}. In: {NASA} Formal
  Methods - Third International Symposium, {NFM} 2011. Proceedings. Lecture
  Notes in Computer Science, vol. 6617, pp. 13--40. Springer (2011)

\bibitem{0002LS07}
Bauer, A., Leucker, M., Schallhart, C.: The good, the bad, and the ugly, but
  how ugly is ugly? In: Sokolsky, O., Tasiran, S. (eds.) Runtime Verification,
  7th International Workshop, {RV} 2007, Vancouver, Canada, March 13, 2007,
  Revised Selected Papers. Lecture Notes in Computer Science, vol. 4839, pp.
  126--138. Springer (2007)

\bibitem{BauerLS10}
Bauer, A., Leucker, M., Schallhart, C.: Comparing {LTL} semantics for runtime
  verification. J. Log. Comput.  20(3),  651--674 (2010)

\bibitem{LTL3Tools}
Bauer, A., Leucker, M., Schallhart, C.: Runtime verification for {LTL} and
  {TLTL}. {ACM} Trans. Softw. Eng. Methodol.  20(4), ~14 (2011)

\bibitem{crowley_context}
Brdiczka, O., Crowley, J.L., Reignier, P.: Learning situation models in a smart
  home. {IEEE} Trans. Systems, Man, and Cybernetics, Part {B}  39(1),  56--63
  (2009)

\bibitem{chen_svm}
Chen, B., Fan, Z., Cao, F.: Activity recognition based on streaming sensor data
  for assisted living in smart homes. In: 2015 International Conference on
  Intelligent Environments, {IE} 2015. pp. 124--127. {IEEE} (2015)

\bibitem{chen_adl}
Chen, L., Hoey, J., Nugent, C.D., Cook, D.J., Yu, Z.: Sensor-based activity
  recognition. {IEEE} Trans. Systems, Man, and Cybernetics, Part {C}  42(6),
  790--808 (2012)

\bibitem{DecentMon}
Colombo, C., Falcone, Y.: Organising {LTL} monitors over distributed systems
  with a global clock. Formal Methods in System Design  49(1-2),  109--158
  (2016)

\bibitem{ex:autosar}
Cotard, S., Faucou, S., B{\'{e}}chennec, J., Queudet, A., Trinquet, Y.: A data
  flow monitoring service based on runtime verification for {AUTOSAR}. In: 14th
  {IEEE} International Conference on High Performance Computing and
  Communication {\&} 9th {IEEE} International Conference on Embedded Software
  and Systems, {HPCC-ICESS} 2012. pp. 1508--1515. {IEEE} Computer Society
  (2012)

\bibitem{crowley_smarthome}
Crowley, J.L., Coutaz, J.: An ecological view of smart home technologies. In:
  De~Ruyter, B., Kameas, A., Chatzimisios, P., Mavrommati, I. (eds.) Ambient
  Intelligence. pp. 1--16. Springer International Publishing, Cham (2015)

\bibitem{orangehome}
Cumin, J., Lefebvre, G., Ramparany, F., Crowley, J.L.: A dataset of routine
  daily activities in an instrumented home. In: Ubiquitous Computing and
  Ambient Intelligence - 11th International Conference, UCAmI 2017,
  Proceedings. Lecture Notes in Computer Science, vol. 10586, pp. 413--425.
  Springer (2017)

\bibitem{LOLA}
D'Angelo, B., Sankaranarayanan, S., S{\'{a}}nchez, C., Robinson, W.,
  Finkbeiner, B., Sipma, H.B., Mehrotra, S., Manna, Z.: {LOLA:} runtime
  monitoring of synchronous systems. In: 12th International Symposium on
  Temporal Representation and Reasoning {(TIME} 2005). pp. 166--174. {IEEE}
  Computer Society (2005)

\bibitem{COEMS}
Decker, N., Dreyer, B., Gottschling, P., Hochberger, C., Lange, A., Leucker,
  M., Scheffel, T., Wegener, S., Weiss, A.: Online analysis of debug trace data
  for embedded systems. In: 2018 Design, Automation {\&} Test in Europe
  Conference {\&} Exhibition, {DATE} 2018. pp. 851--856. {IEEE} (2018)

\bibitem{artifact}
El-Hokayem, A., Falcone, Y.: {THEMIS Smart Home Artifact Repository},
  \url{gitlab.inria.fr/monitoring/themis-rv18smarthome}

\bibitem{themis_issta}
El{-}Hokayem, A., Falcone, Y.: Monitoring decentralized specifications. In:
  Antoine El{-}Hokayem and Yli{\`{e}}s Falcone  \cite{DBLP:conf/issta/2017},
  pp. 125--135

\bibitem{themis_tool}
El{-}Hokayem, A., Falcone, Y.: {THEMIS:} a tool for decentralized monitoring
  algorithms. In: Antoine El{-}Hokayem and Yli{\`{e}}s Falcone
  \cite{DBLP:conf/issta/2017}, pp. 372--375

\bibitem{El-HokayemF18a}
El{-}Hokayem, A., Falcone, Y.: Bringing runtime verification home. In: Colombo,
  C., Leucker, M. (eds.) Runtime Verification - 18th International Conference,
  {RV} 2018, Limassol, Cyprus, November 10-13, 2018, Proceedings. Lecture Notes
  in Computer Science, vol. 11237, pp. 222--240. Springer (2018),
  \url{https://doi.org/10.1007/978-3-030-03769-7}

\bibitem{tosem}
El{-}Hokayem, A., Falcone, Y.: On the monitoring of decentralized
  specifications. ACM Transactions on Software Engineering and Methodology
  (TOSEM)  (2019), to appear.

\bibitem{Falcone10}
Falcone, Y.: You should better enforce than verify. In: Barringer, H., Falcone,
  Y., Finkbeiner, B., Havelund, K., Lee, I., Pace, G.J., Rosu, G., Sokolsky,
  O., Tillmann, N. (eds.) Runtime Verification - First International
  Conference, {RV} 2010, St. Julians, Malta, November 1-4, 2010. Proceedings.
  Lecture Notes in Computer Science, vol. 6418, pp. 89--105. Springer (2010)

\bibitem{RVTutorial}
Falcone, Y., Havelund, K., Reger, G.: A tutorial on runtime verification. In:
  Engineering Dependable Software Systems, {NATO} science for peace and
  security series, {d:} information and communication security, vol.~34, pp.
  141--175. {ios} press (2013)

\bibitem{FalconeJMP16}
Falcone, Y., J{\'{e}}ron, T., Marchand, H., Pinisetty, S.: Runtime enforcement
  of regular timed properties by suppressing and delaying events. Sci. Comput.
  Program.  123,  2--41 (2016)

\bibitem{FalconeMRS18}
Falcone, Y., Mariani, L., Rollet, A., Saha, S.: Runtime failure prevention and
  reaction. In: Bartocci and Falcone  \cite{lncs/10457}, pp. 103--134

\bibitem{FalconeMFR11}
Falcone, Y., Mounier, L., Fernandez, J., Richier, J.: Runtime enforcement
  monitors: composition, synthesis, and enforcement abilities. Formal Methods
  in System Design  38(3),  223--262 (2011)

\bibitem{Haghighi:2015}
Haghighi, I., Jones, A., Kong, Z., Bartocci, E., Gros, R., Belta, C.: Spatel: A
  novel spatial-temporal logic and its applications to networked systems. In:
  Proceedings of the 18th International Conference on Hybrid Systems:
  Computation and Control. pp. 189--198. HSCC '15, ACM, New York, NY, USA
  (2015)

\bibitem{ADLBEEPBEEP}
Hall{\'{e}}, S., Gaboury, S., Bouchard, B.: Activity recognition through
  complex event processing: First findings. In: Artificial Intelligence Applied
  to Assistive Technologies and Smart Environments, Papers from the 2016 {AAAI}
  Workshop. {AAAI} Workshops, vol. {WS-16-01}. {AAAI} Press (2016)

\bibitem{HavelundG05}
Havelund, K., Goldberg, A.: Verify your runs. In: Meyer, B., Woodcock, J.
  (eds.) Verified Software: Theories, Tools, Experiments, First {IFIP} {TC}
  2/WG 2.3 Conference, {VSTTE} 2005, Zurich, Switzerland, October 10-13, 2005,
  Revised Selected Papers and Discussions. Lecture Notes in Computer Science,
  vol. 4171, pp. 374--383. Springer (2005)

\bibitem{LamaConv}
{Institute for Software Engineering and Programming Languages}: {LamaConv} -
  {Logics and Automata Converter Library},
  \url{www.isp.uni-luebeck.de/lamaconv}

\bibitem{KasterenEK10}
van Kasteren, T., Englebienne, G., Kr{\"{o}}se, B.J.A.: Transferring knowledge
  of activity recognition across sensor networks. In: Pervasive Computing, 8th
  International Conference, Pervasive 2010. Proceedings. Lecture Notes in
  Computer Science, vol. 6030, pp. 283--300. Springer (2010)

\bibitem{Katz-ADL}
Katz, S.: Assessing self-maintenance: Activities of daily living, mobility, and
  instrumental activities of daily living. Journal of the American Geriatrics
  Society  31(12),  721--727 (1983)

\bibitem{Paula}
Lago, P., Lang, F., Roncancio, C., Jim{\'{e}}nez{-}Guar{\'{\i}}n, C., Mateescu,
  R., Bonnefond, N.: The {ContextAct@A4H} real-life dataset of daily-living
  activities - activity recognition using model checking. In: Modeling and
  Using Context - 10th International and Interdisciplinary Conference,
  {CONTEXT} 2017, Proceedings. Lecture Notes in Computer Science, vol. 10257,
  pp. 175--188. Springer (2017)

\bibitem{jlp/LeuckerS09}
Leucker, M., Schallhart, C.: A brief account of runtime verification. J. Log.
  Algebr. Program.  78(5),  293--303 (2009)

\bibitem{ex:medical}
Leucker, M., Schmitz, M., {\`{a}}~Tellinghusen, D.: Runtime verification for
  interconnected medical devices. In: Leveraging Applications of Formal
  Methods, Verification and Validation: Discussion, Dissemination, Applications
  - 7th International Symposium, ISoLA 2016, Proceedings, Part {II}. Lecture
  Notes in Computer Science, vol. 9953, pp. 380--387 (2016)

\bibitem{home_elderly}
Majumder, S., Aghayi, E., Noferesti, M., Memarzadeh{-}Tehran, H., Mondal, T.,
  Pang, Z., Deen, M.J.: Smart homes for elderly healthcare - recent advances
  and research challenges. Sensors  17(11),  2496 (2017)

\bibitem{MCL}
Mateescu, R., Thivolle, D.: A model checking language for concurrent
  value-passing systems. In: {FM} 2008: Formal Methods, 15th International
  Symposium on Formal Methods, Proceedings. Lecture Notes in Computer Science,
  vol. 5014, pp. 148--164. Springer (2008)

\bibitem{Pnueli77}
Pnueli, A.: The temporal logic of programs. In: 18th Annual Symposium on
  Foundations of Computer Science, Providence, Rhode Island, USA, 31 October -
  1 November 1977. pp. 46--57. {IEEE} Computer Society (1977)

\bibitem{CRDT}
Shapiro, M., Pregui{\c{c}}a, N.M., Baquero, C., Zawirski, M.: Conflict-free
  replicated data types. In: D{\'{e}}fago, X., Petit, F., Villain, V. (eds.)
  Stabilization, Safety, and Security of Distributed Systems - 13th
  International Symposium, {SSS} 2011, Grenoble, France, October 10-12, 2011.
  Proceedings. Lecture Notes in Computer Science, vol. 6976, pp. 386--400.
  Springer (2011)

\bibitem{CBSE}
Szyperski, C.A., Gruntz, D., Murer, S.: Component software - beyond
  object-oriented programming, 2nd Edition. Addison-Wesley component software
  series, Addison-Wesley (2002)

\bibitem{tapia_adl}
Tapia, E.M., Intille, S.S., Larson, K.: Activity recognition in the home using
  simple and ubiquitous sensors. In: Pervasive Computing, Second International
  Conference, {PERVASIVE} 2004, Vienna, Austria, April 21-23, 2004,
  Proceedings. Lecture Notes in Computer Science, vol. 3001, pp. 158--175.
  Springer (2004)

\bibitem{home_impaired}
Thapliyal, H., Nath, R.K., Mohanty, S.P.: Smart home environment for mild
  cognitive impairment population: Solutions to improve care and quality of
  life. {IEEE} Consumer Electronics Magazine  7(1),  68--76 (2018)

\end{thebibliography}
